\documentclass[aps,prl,twocolumn,superscriptaddress,footinbib,notitlepage,floatfix]{revtex4-2}
\usepackage[latin9]{inputenc}
\usepackage{color}
\usepackage{amsmath}
\usepackage{multirow}
\usepackage{bbm}
\usepackage{amssymb}
\usepackage{makecell}
\usepackage{float}
\usepackage{lineno}
\usepackage{booktabs}
\usepackage{graphicx}
\usepackage[unicode=true,bookmarks=true,bookmarksnumbered=false,bookmarksopen=false,breaklinks=false,pdfborder={0 0 1},backref=false,colorlinks=true]{hyperref}
\hypersetup{colorlinks=true, linkcolor=blue, urlcolor=blue, citecolor=blue, pdfstartview={FitH}, hyperfootnotes=false, unicode=true}
\makeatletter
\renewcommand{\arraystretch}{1.35}
\usepackage[table,xcdraw]{xcolor}
\usepackage{framed}
\usepackage{fbox}
\usepackage{amsfonts}
\usepackage{tabularx}
\usepackage{dcolumn}
\usepackage{bm}
\usepackage{epstopdf}
\usepackage{times}
\hypersetup{urlcolor=blue}

\usepackage{tikz}
\usetikzlibrary{quantikz}
\usepackage{algorithm}
\usepackage{algorithmic}
\usepackage{amsthm}

\newtheorem{proposition}{Proposition}

\newtheorem{lemma}{Lemma}

\definecolor{blue}{rgb}{0,0,1}
\definecolor{red}{rgb}{1,0,0}
\definecolor{green}{rgb}{0,1,0}

\renewcommand{\figurename}{Fig.}
\renewcommand{\tablename}{Tab.}

\begin{document}

\title{Quantum error correction at ultra-low overhead}

\author{Zhide Lu}
\affiliation{Shanghai Qi Zhi Institute, Shanghai 200232, China}

\author{Weikang Li}
\email{weikang\_li@mail.tsinghua.edu.cn}
\affiliation{Center for Quantum Information, IIIS, Tsinghua University, Beijing 100084, China}

\author{Dong-Ling Deng}
\email{dldeng@tsinghua.edu.cn}
\affiliation{Center for Quantum Information, IIIS, Tsinghua University, Beijing 100084, China}
\affiliation{Shanghai Qi Zhi Institute, Shanghai 200232, China}
\affiliation{Hefei National Laboratory, Hefei 230088, China}

\begin{abstract}\noindent
\textbf{\bfseries\boldmath
Suppressing errors is the central challenge for useful large-scale quantum computing.  While quantum error correction promises a viable solution to this challenge, existing codes typically suffer from trade-offs among encoding efficiency, error threshold, and hardware feasibility. Here, we introduce Cornucopia codes, a family of practical, hardware-efficient quantum low-density parity-check codes that achieve an ultra-high encoding rate exceeding $1/2$ while maintaining a pseudo-threshold exceeding $0.4\%$ under the standard circuit-level noise model. Inspired by recent affine-permutation-based code constructions and the long-range connectivity available in reconfigurable neutral-atom arrays, we adopt a structured code geometry in which the code layout, atom rearrangement, and syndrome extraction schedule are co-designed. This structure enables nonlocal syndrome measurements through simple, parallel atom rearrangements. 
A complete syndrome extraction cycle measures all $X$- and $Z$-type checks in parallel with $12$ entangling layers, independent of the code size.
The resulting threshold is comparable to those of the surface code and bivariate bicycle codes. In particular, a single code block $[[2844,1426,18]]$ encodes $1{,}426$ distance-$18$ logical qubits, achieving an extrapolated logical error rate of $2.6\times10^{-16}$ ($1.9\times10^{-31}$) per logical qubit per cycle, assuming the physical error rate of $0.1\%$ ($0.01\%$). By comparison, a bivariate bicycle code implementation would require more than $68{,}000$ physical qubits to encode the same number of logical qubits at a comparable logical error rate. These results bring demonstrations of ultra-low-overhead quantum error correction within the reach of near-term quantum processors. 
}
\end{abstract}

\maketitle

\noindent 
Quantum computers hold the potential to surpass classical computers in a wide range of applications, including factoring~\cite{Shor1994Algorithms}, chemistry~\cite{AspuruGuzik2005Simulated,Lanyon2010Towards,Kandala2017Hardware}, finance~\cite{Herman2023Quantum}, and artificial intelligence~\cite{Biamonte2017Quantum,Havlicek2019Supervised,Saggio2021Experimental}. However, the fragility of quantum information and the error-prone nature of quantum operations make the construction of large-scale, fault-tolerant quantum computers a formidable challenge. To realize a practical quantum advantage, errors at the physical level must be suppressed sufficiently to execute large and deep quantum circuits with the desired accuracy. Quantum error correction provides a route to fault tolerance by redundantly encoding logical qubits into a larger number of physical qubits, allowing errors to be diagnosed and corrected through repeated syndrome measurements~\cite{Gottesman1997Stabilizer,Dennis2002Topological}. Experimental demonstrations of quantum error correction have been reported across diverse platforms, including trapped ions~\cite{Paetznick2024Demonstration,Daguerre2025Experimentala,Dasu2026Computing,Tham2026Breakeven}, superconducting processors~\cite{GoogleQuantumAI2023Suppressing,Ni2023Beating,Gupta2024Encoding,Lacroix2025Scaling,GoogleQuantumAIandCollaborators2025Quantum,He2025Experimental,Rosenfeld2025Magic,Wang2026Demonstration} and neutral atoms~\cite{Bluvstein2024Logical,Reichardt2024Logical,SalesRodriguez2025Experimental,Bluvstein2026Faulttolerant}. Despite this rapid progress, the substantial physical-resource overhead required to protect logical information remains a major obstacle to scalable quantum computation.

The surface code has emerged as a leading approach because of its high error threshold and compatibility with geometrically local operations~\cite{Kitaev2003Faulttolerant,Bravyi1998Quantum}. These advantages, however, come at the cost of a low encoding rate: the number of physical qubits required per logical qubit grows rapidly with code distance. This overhead becomes particularly severe for algorithms requiring hundreds or thousands of logical qubits, for which the physical-qubit requirements can far exceed the capabilities of current processors. Reducing this overhead without sacrificing fault tolerance and hardware feasibility is therefore of crucial importance.  More general quantum low-density parity-check (qLDPC) codes offer a promising way out~\cite{Breuckmann2021Quantum,Zhou2025Opportunities}. By encoding multiple logical qubits into a single code block using bounded-weight parity checks, these qLDPC codes can achieve substantially higher encoding rates than the surface code. 
Recent theoretical constructions have produced code families with increasingly favourable rates and distances~\cite{Bravyi2024Highthreshold,Liang2025Generalized,Liang2025Planar,Liu2026LargeLanguageModel,Choe2026Barbell,Hong2026Quantum,Nixon2026Vine,Gu2026Nearestneighbour,Aydin2026Breakinga,Khesin2026Mirror,Lee2026Logical,Bhardwaj2026Highrate,Zheng2026Logical}. In particular, bivariate bicycle codes and related proposals have demonstrated the potential for substantial reductions in physical-qubit overhead, with encoding rates typically around $1/10$~\cite{Bravyi2024Highthreshold}. 
Yet favourable code parameters alone do not guarantee a practical implementation. Nonlocal parity checks may demand complex connectivity and deeply serialized syndrome extraction circuits. These operations can introduce additional errors and increase idle time, diminishing the theoretical advantages of such codes. Conversely, codes designed around strictly local hardware constraints retain a high threshold but encode logical information inefficiently. Existing approaches thus generally face a three-way trade-off among encoding rate, error threshold and hardware feasibility.

\begin{figure*}[t]
\centering
\includegraphics[width=0.98\textwidth]{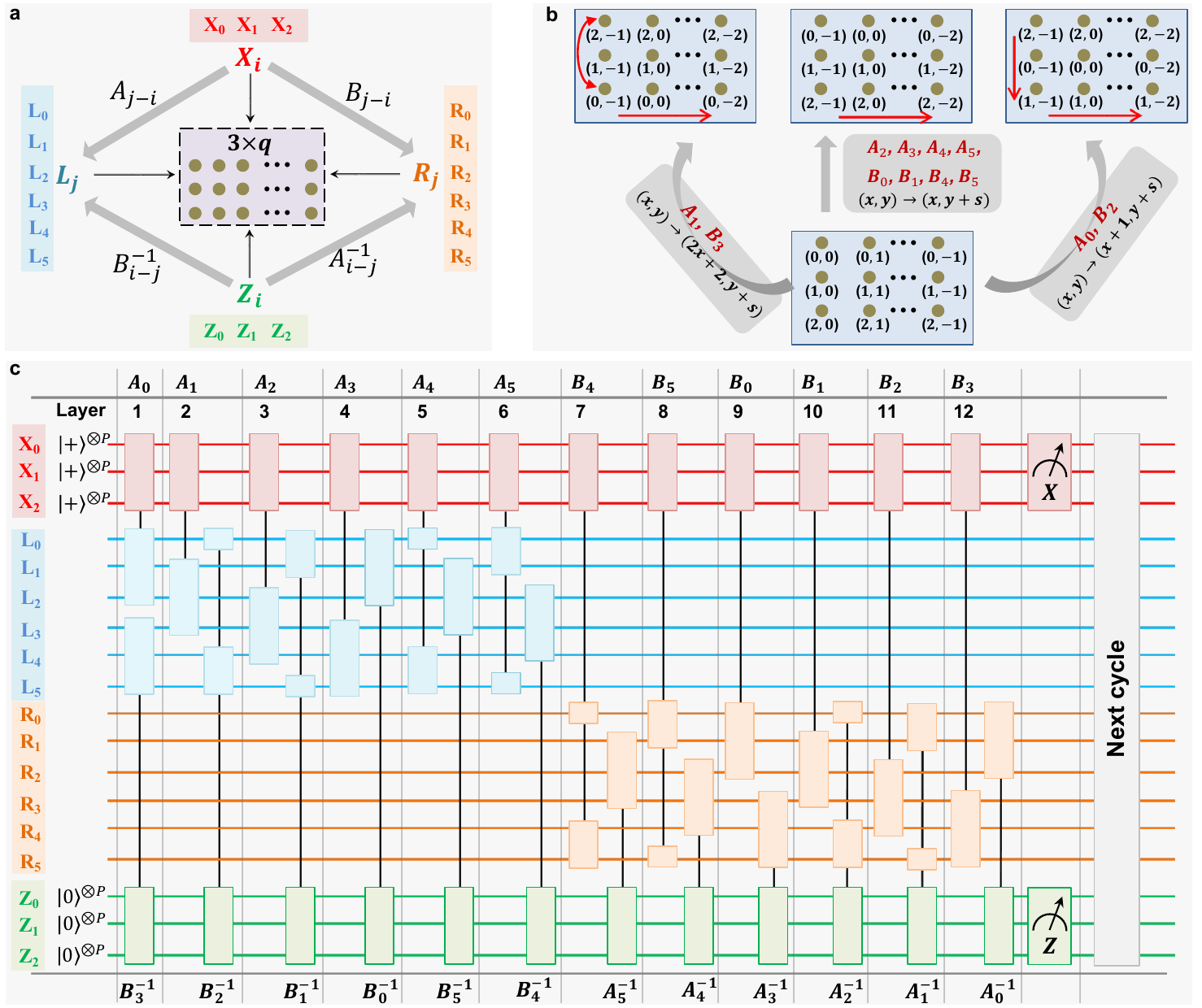}
\caption{\textbf{Construction and implementation of Cornucopia codes.}
\textbf{a,} Block structure and connectivity. Data qubits are partitioned into six $L$-type and six $R$-type blocks, while check qubits comprise three $X$-type and three $Z$-type blocks, with individual qubits indexed by row-column coordinates $(x,y) \in \mathbb{Z}_3 \times \mathbb Z_q$. The connectivity between check and data blocks is governed by operators $A_k$ and $B_k$ (indices evaluated modulo $6$), each defining a coordinate permutation that specifies a one-to-one physical coupling pattern. Specifically, $X_i$ check blocks couple to $L_j$ and $R_j$ via $A_{j-i}$ and $B_{j-i}$, whereas $Z_i$ check blocks couple via $B^{-1}_{i-j}$ and $A^{-1}_{i-j}$, respectively.
\textbf{b,} Geometric action of coordinate permutations. The permutations are classified into three families acting on the row-column grid. The operators $A_1$ and $B_3$ implement the affine transformation $(x, y) \mapsto (2x+2, y + s)$, which inverts the row order by swapping rows $0$ and $2$ while cyclically shifting columns; $A_0$ and $B_2$ define a translation $(x, y) \mapsto (x+1, y + s)$ that cyclically shifts both rows and columns. The remaining operators act as pure column translations $(x, y) \mapsto (x, y + s)$ that leave the row indices invariant. The example shown uses $s=1$. 
\textbf{c,} Circuit scheduling for syndrome extraction. A full syndrome measurement cycle is executed across twelve sequential layers of transversal CNOT gates. For these CNOT gates, the $X$-type and $Z$-type check blocks serve as the controls and targets, respectively. Within each layer, check blocks couple to data blocks, with the exact physical qubit coordinate pairings dictated by $A_i, B_i$ and their inverse $A_i^{-1}, B_i^{-1}$. See Extended Data Tab.~\ref{tab:depth12_syndrome_schedule} for details.
}
\label{fig:1}
\end{figure*}

\begin{table}[t]
\caption{\textbf{Representative instances of Cornucopia codes.}
All code instances have weight-$12$ check operators and a depth-$12$ syndrome measurement circuit.
The encoding rate $r=k/n$ is defined for the data qubits and does not include the ancillary check qubits used for syndrome extraction.
The code distance $d$ is exactly certified by exhaustively excluding all candidate logical operators of weight less than $d$ and identifying an explicit logical operator of weight $d$. 
The circuit-level distance $d_{\mathrm{circ}}$ is defined as the minimum number of independent physical faults that produce an undetectable logical error.
The logical error rates per logical qubit per cycle, $p_{\mathrm{L}}$, at physical error rates of $p=10^{-3}$ and $10^{-4}$ are obtained from the circuit-level simulations shown in Fig.~\ref{fig:2}.
}
\label{tab:family}
\centering
\setlength{\tabcolsep}{4pt}
\begin{tabular}{c c c c c}
\toprule
{\bfseries\boldmath$[[n,k,d]]$}
& {\bfseries\boldmath$r$}
& {\bfseries\boldmath$d_{\rm cir}$}
& {\bfseries\boldmath$p_{L}(10^{-3})$}
& {\bfseries\boldmath$p_{L}(10^{-4})$}\\
\midrule
$[[252,130,6]]$    & $0.516$ & $\leq 5$  & $3.3\times10^{-6}$  & $1.9\times10^{-9}$  \\
$[[576,292,8]]$    & $0.507$ & $\leq 8$  & $7.0\times10^{-9}$  & $7.8\times10^{-13}$ \\
$[[900,454,10]]$   & $0.504$ & $\leq 10$ & $4.1\times10^{-10}$ & $2.4\times10^{-14}$ \\
$[[1044,526,12]]$  & $0.504$ & $\leq 11$ & $2.0\times10^{-12}$ & $6.3\times10^{-21}$ \\
$[[1764,886,14]]$  & $0.502$ & $\leq 14$ & $5.6\times10^{-13}$   & $2.7\times10^{-21}$                   \\
$[[2304,1156,16]]$ & $0.502$ & $\leq 16$ & $9.6\times10^{-15}$ & $4.3\times10^{-27}$ \\
$[[2844,1426,18]]$ & $0.501$ & $\leq 18$ & $2.6\times10^{-16}$ & $1.9\times10^{-31}$ \\
\bottomrule
\end{tabular}
\end{table}

Reconfigurable neutral-atom arrays offer a compelling platform to overcome this bottleneck. Their long-range and reconfigurable connectivity~\cite{Chiu2025Continuous,Manetsch2025Tweezer,Lin2026Sustaining,Evered2026Highfidelity} enable complex interactions that remain difficult on architectures with fixed nearest-neighbor couplings. Nevertheless, directly embedding an arbitrary qLDPC code onto such a processor can require irregular or highly serialized atom rearrangements, resulting in prolonged syndrome extraction times and increased vulnerability to decoherence. Fully exploiting the capabilities of neutral-atom hardware therefore requires moving beyond the mere adaptation of existing codes to the device.
Instead, the code structure, physical layout, atom-rearrangement protocol and syndrome extraction circuits should be designed together.

Here, building on recent ultra-high-rate code constructions~\cite{Kasai2026Breaking,Zhao2026UltraHighRate}, we introduce Cornucopia codes, a family of qLDPC codes constructed in this hardware-code co-design principle. 
Cornucopia codes combine three key advantages. First, they achieve an ultra-high code rate exceeding $1/2$, encoding more than one logical qubit for every two physical data qubits and sharply reducing the resource cost of fault-tolerant computation. 
Second, despite this high rate, they retain a pseudo-threshold exceeding $0.4\%$ under the standard circuit-level noise model---a practically promising regime comparable to those reported for surface codes and bivariate bicycle codes.
Third, they possess a regular geometry tailored to reconfigurable neutral-atom arrays. This structure reduces nonlocal syndrome measurements to simple, parallel rigid qubit translations, and ensures a constant-depth syndrome extraction circuit regardless of the code length.
The estimated total qubit-routing time per syndrome extraction cycle is orders of magnitude shorter than the coherence times already demonstrated in neutral-atom experiments, further underscoring the feasibility of implementing Cornucopia codes on existing neutral-atom platforms.
In addition, Cornucopia codes have native structural symmetries necessary for efficient logical routing.

\vspace{.2cm}
\noindent\textbf{\large{}Cornucopia code construction} \\
\noindent 
The Cornucopia code family represents a class of $[[n, k, d]]$ Calderbank-Shor-Steane (CSS)  stabilizer codes, encoding $k$ logical qubits into $n$ physical data qubits with a minimum distance $d$. As illustrated in Fig.~\ref{fig:1}\textbf{a}, the spatial layout partitions the $n$ data qubits into two registers. Each register comprises six distinct data blocks, labeled $L_i$ and $R_i$ for $i \in \{0, 1, \dots, 5\}$. To perform syndrome extraction, the code introduces $n/2$ ancilla qubits, which are organized into three $X$-type ($X_i$) and three $Z$-type ($Z_i$) check blocks for $i \in \{0, 1, 2\}$.
Geometrically, each data or check block contains $P=3q$ qubits arranged as a $3 \times q$ rectangular array.
This spatial layout yields a convenient geometric description of the checks. 
A single check operator is specified by the twelve physical coordinates of its target qubits across the twelve data blocks.
These twelve coordinates are determined by a set of coordinate permutations acting on the $3 \times q$ grid.
Specifically, for an $X$-type check qubit located at coordinate $(x, y)$ within block $X_i$, its coupled data qubit in block $L_j$ is situated at the permuted coordinate $A_{j-i}(x, y)$, while its counterpart in block $R_j$ is located at $B_{j-i}(x, y)$, with all block indices evaluated modulo $6$; a $Z$-type check qubit at $(x, y)$ within block $Z_i$ locates its target data qubits in $L_j$ and $R_j$ at $B_{i-j}^{T}(x, y)$ and $A_{i-j}^{T}(x, y)$, respectively (Methods). 
Crucially, as shown in Fig.~\ref{fig:1}\textbf{b}, all coordinate permutations $A_i$ and $B_i$ decompose into collective cyclic shifts along the column axis, with nontrivial permutations restricted strictly to the row axis. 
This property is deliberately designed  to suit the hardware capabilities of reconfigurable neutral-atom arrays, as discussed below.

We present several concrete instances of Cornucopia codes in Tab.~\ref{tab:family}.
By construction, all presented instances achieve a high encoding rate of $r>1/2$. This lower bound arises naturally from the fact: since the total number of check qubits is exactly $n/2$, the number of encoded logical qubits is guaranteed to be at least $n/2$.
By contrast, bivariate bicycle code instances such as $[[144, 12, 12]]$ and $[[288, 12, 18]]$ yield significantly lower encoding rates of $1/12$ and $1/24$, respectively. All code distances reported in Tab.~\ref{tab:family} are exact and have been rigorously certified (Methods).

To prevent error accumulation, one must periodically measure stabilizers and perform error correction.
We integrate all individual stabilizer measurements into a syndrome measurement circuit, as shown in Fig.~\ref{fig:1}\textbf{c}. The full syndrome cycle begins with the initialization of all check qubits. 
The circuit then executes twelve layers of non-overlapping CNOT gates---an entangling depth that remains strictly independent of the code length. 
Within each layer, every check block is coupled to its target data block via transversal CNOT gates. Between consecutive layers, the check qubits are physically rearranged to bring paired data and check qubits into interaction range. Finally, the circuit measures all check qubits to extract the error syndromes.

\begin{figure}[t]
\centering
\includegraphics[width=0.48\textwidth]{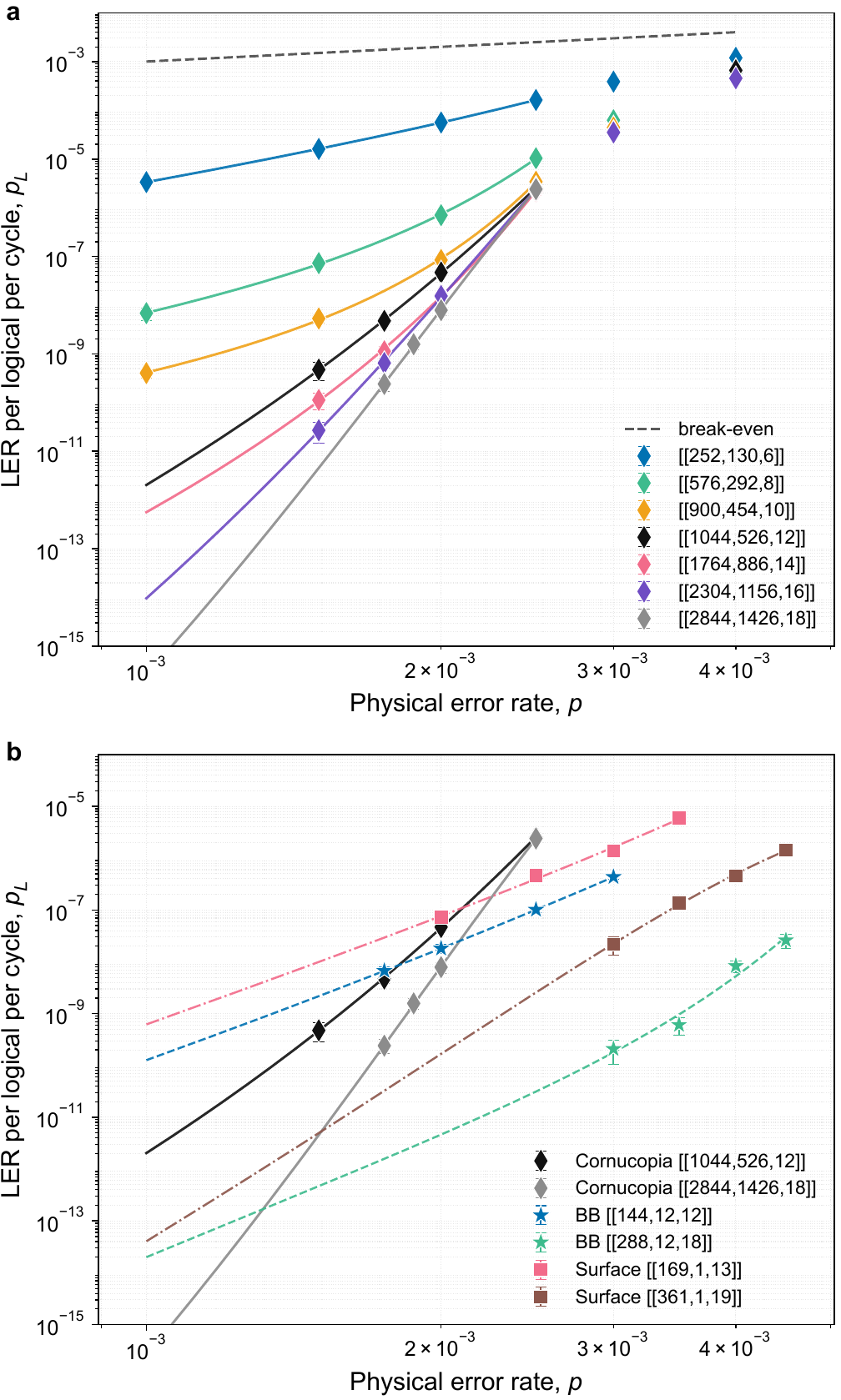}
\caption{\textbf{Logical performance of Cornucopia codes.}
\textbf{a}, Logical error rate per logical qubit per cycle, $p_{\mathrm{L}}$, as a function of the physical error rate, $p$, for the Cornucopia codes in Tab.~\ref{tab:family}. The grey dashed line indicates the break-even condition $p_{\mathrm{L}} = p$.
\textbf{b}, Comparison of the Cornucopia codes $[[1044, 526, 12]]$ and $[[2844, 1426, 18]]$ against representative bivariate bicycle codes~\cite{Bravyi2024Highthreshold} and rotated surface codes. 
The surface code encodes a single logical qubit in an independent patch.
In both panels, markers denote numerical simulation results, with error bars representing the $1\sigma$ statistical uncertainty. The solid and dash-dotted lines represent extrapolations to the low-error regime obtained via the fitting ansatz~(Methods). LER, logical error rate. BB, bivariate bicycle code.}
\label{fig:2}
\end{figure}

\begin{figure*}[t]
\centering
\includegraphics[width=0.99\textwidth]{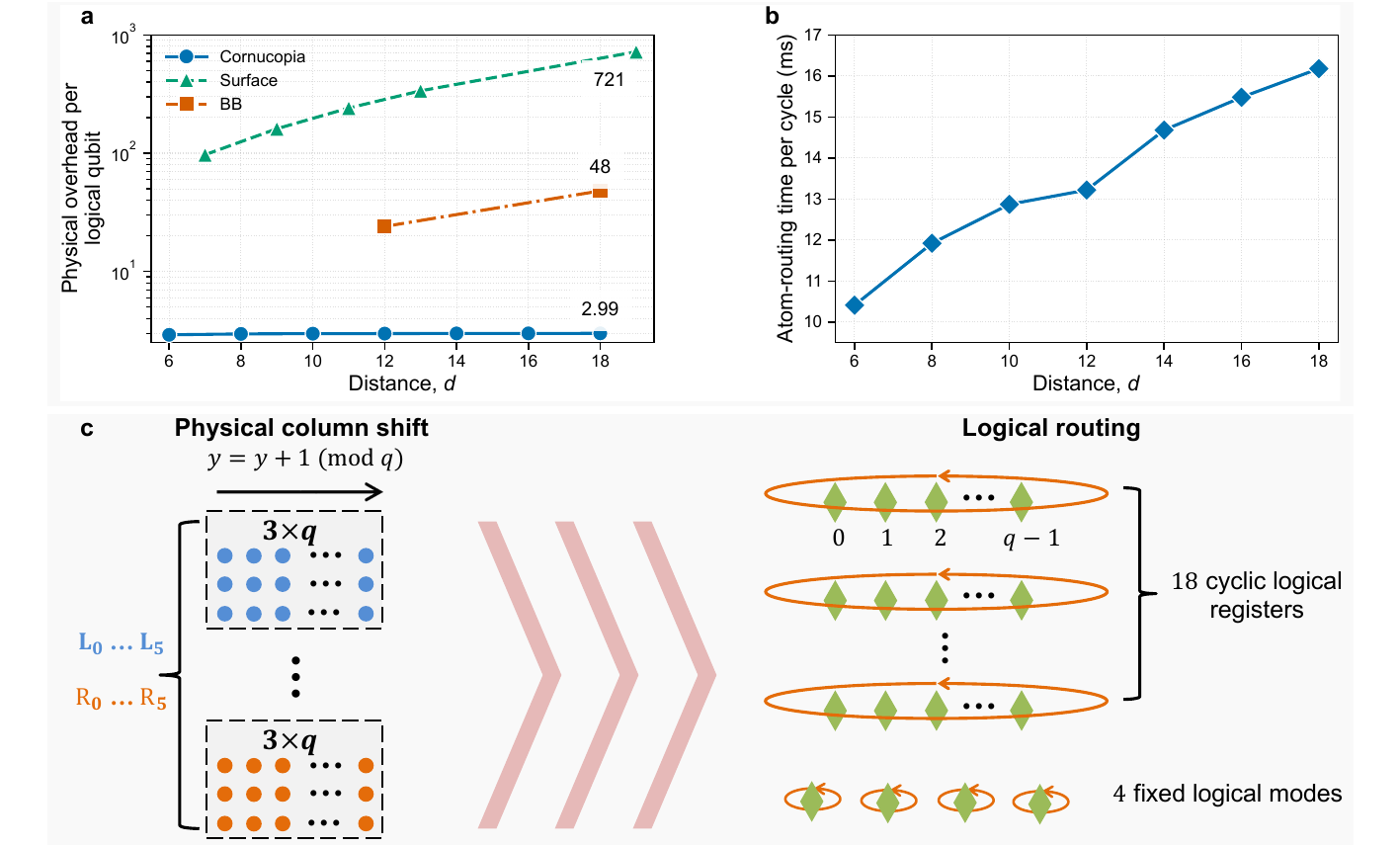}
\caption{\textbf{Hardware overhead and logical routing of Cornucopia codes.}
\textbf{a}, Physical overhead per logical qubit as a function of the code distance $d$. The physical overhead takes into account both data qubits and check qubits required for syndrome extraction.
Cornucopia codes (blue) maintain a low overhead of ${\sim}3$ across all distances. 
In contrast, the per-logical physical overhead for surface codes (green) and bivariate bicycle codes (orange) scales unfavorably, reaching $721$ (at $d=19$) and $48$ (at $d=18$), respectively. 
\textbf{b}, Qubit-routing time per syndrome extraction cycle as a function of the code distance $d$ for Cornucopia codes. The physical execution time exhibits a modest increase from $10.4$ ms to $16$ ms as the code scales from $d=6$ to $d=18$.
\textbf{c}, Illustration of the logical routing induced by code automorphisms. A global physical column shift, $y \mapsto y + 1 \pmod{q}$, applied across all twelve $3 \times q$ data blocks, directly induces a macroscopic permutation on the encoded logical space. 
As detailed in Methods, this operation decomposes into parallel cyclic shifts across $18$ logical registers of length $q$, accompanied by $4$ invariant logical modes, providing an efficient logical routing primitive.
}
\label{fig:3}
\end{figure*}

\vspace{.2cm}
\noindent\textbf{\large{}Decoding and logical performance} \\
We evaluate the logical performance of Cornucopia codes using the standard circuit-level noise model. Each physical operation within the syndrome measurement circuit, including CNOT gates, data qubit initializations, reset of check qubits, and measurements, fails independently with a uniform physical error rate $p$. 
Given the long coherence times inherent to neutral-atom platforms, idling errors are neglected. 
For each code instance in Tab.~\ref{tab:family}, we simulate $N_c$ syndrome extraction cycles, followed by a transversal destructive measurement of all data qubits in both logical $Z$ and $X$ bases. 
The obtained error syndromes are then decoded by a classical decoding algorithm. In particular, we use relay-BP (belief propagation)~\cite{Muller2025Improved,Maurer2025Realtime} as the primary decoder. When relay-BP fails to converge within the prescribed iteration limit, we invoke BP with ordered statistics decoding (BP-OSD)~\cite{Roffe2020Decoding,Panteleev2021Degenerate} as a fallback. 
This hybrid strategy leverages the high efficiency of relay-BP to rapidly resolve typical error syndromes, reserving the heavier OSD post-processing for the most stubborn error instances (Methods).
Decoding is deemed successful if and only if the product of the inferred correction and the actual accumulated error is a stabilizer.

Let $P_{\mathrm{fail}}$ denote the probability that at least one of the $k$ encoded logical qubits fails after $N_c$ syndrome extraction cycles, averaged over the logical $Z$ and $X$ bases. We define the effective logical error rate per logical qubit per cycle as $p_{\mathrm{L}} = 1 - (1 - P_{\mathrm{fail}})^{1/(kN_c)}$. The data points presented in Fig.~\ref{fig:2} correspond to these numerically simulated logical error rates, which are subsequently extrapolated to lower physical error regimes (Methods). 
As shown in Fig.~\ref{fig:2}\textbf{a}, all simulated Cornucopia codes exhibit a pseudo-threshold (defined by the break-even condition $p_{\mathrm{L}}=p$) exceeding $0.4\%$. 
At a physical error rate of $p=10^{-3}$, a target within reach of near-term neutral-atom experiments~\cite{Evered2026Highfidelity}, a single distance-$18$ Cornucopia code block $[[2844,1426,18]]$ encodes $1426$ logical qubits, achieving an extrapolated logical error rate of $2.6 \times 10^{-16}$. The total number of physical qubits required for this encoding is $4266$, including $2844$ data qubits and $1422$ check qubits.
For comparison, encoding $1426$ logical qubits into separate patches of the bivariate bicycle code $[[288,12,18]]$ requires $\lceil 1426/12 \rceil = 119$ independent blocks, demanding an overhead of more than $68,000$ physical qubits. Thus, the distance-$18$ Cornucopia code offers a greater than tenfold reduction in physical qubit overhead compared to the bivariate bicycle code in the experimentally relevant error regime.

\vspace{2mm}
\noindent\textbf{\large{}Implementation in neutral-atom arrays} \\
To facilitate the implementation of Cornucopia codes in reconfigurable neutral-atom arrays, qubit rearrangement is required to bring paired data and check qubits into interaction range between consecutive layers of the syndrome measurement circuit---a process that dominates the syndrome cycle time (Methods). 
Crucially, all required coordinate permutations (Fig.~\ref{fig:1}\textbf{b}) decompose into collective cyclic shifts along the column axis, with nontrivial permutations restricted to the row axis. These collective shifts compile directly into acousto-optic deflector (AOD)-compatible global rigid motions, thereby minimizing atom-routing overhead. 
This property allows the non-local syndrome extraction of Cornucopia codes to be efficiently translated into a deterministic choreography of collective cyclic shifts and a few parallel row swaps, enabling a hardware-efficient implementation.

As illustrated in Extended Data Figs.~\ref{fig:atom_moving} and \ref{fig:atom_moving2}, a Cornucopia code is physically implemented by vertically stacking the twelve data blocks and six check blocks into a global two-dimensional array. Between consecutive layers of syndrome extraction, the data blocks are shifted as intact units to align with their target check blocks, while the check blocks undergo internal coordinate permutations.
The total qubit rearrangement time for a full syndrome cycle naturally partitions into two distinct contributions: data-block routing and check-block routing.
The data-block routing involves macroscopic vertical cyclic shifts and the global swapping of the $L$- and $R$-type data block arrays. 
Because these operations traverse a fixed spatial distance, they incur an architectural constant-time overhead (estimated at $6.07\,\text{ms}$) that remains invariant regardless of the code size.
In contrast, the check-block routing requires collective cyclic shifts along the column axis. Given that the maximum required shift distance grows linearly with the column dimension $q$, this routing time inherently scales as $O(\sqrt{q})$, or equivalently, $O(\sqrt{n})$.

We estimate the total qubit rearrangement time for a full syndrome cycle across all Cornucopia codes in Tab.~\ref{tab:family}, with the results provided in Fig.~\ref{fig:3}\textbf{b} and Extended Data Tab.~\ref{tab:movement_cost}. Strikingly, even for the $[[2844, 1426, 18]]$ code, the total qubit-routing time per syndrome cycle is merely $16.18$\,ms, three orders of magnitude shorter than the ${\sim}10$\,s coherence times achieved in neutral-atom experiments~\cite{Jenkins2022Ytterbium,Ma2022Universal}. These estimates indicate that the hardware implementation of Cornucopia codes is well within the reach of current experimental capabilities.

In addition, efficient logical routing--the transport of logical observables to interaction zones or lattice surgery boundaries--is essential for scalable fault-tolerant quantum computation. 
Cornucopia codes provide a native routing mechanism by leveraging their intrinsic column-shift automorphism (Methods). 
As shown in Fig.~\ref{fig:3}\textbf{c}, a global physical column shift $y\mapsto y+1 \pmod q$ across all data blocks acts as a macroscopic permutation on the encoded logical space.
Within a symmetry-adapted basis, the logical Pauli observables (e.g., the logical $Z$) can be partitioned into $18$ independent cyclic registers, each containing $q$ logical operators, supplemented by four invariant logical modes. Thus, a single physical column shift functions as a highly parallelized operation, simultaneously executing a cyclic translation of logical operators within each of these $18$ logical registers.

\vspace{0.2cm}
\noindent\textbf{\large{}Discussion} \\
In summary, we have constructed a family of qLDPC codes with encoding rates exceeding $1/2$ that can be efficiently implemented on reconfigurable neutral-atom arrays. Numerical simulations under a standard circuit-level noise model show that the Cornucopia codes compare favorably with both surface codes and bivariate bicycle codes. In the practically relevant regime of physical error rates around $10^{-3}$, they achieve comparable logical-error suppression while reducing the physical-qubit overhead by more than an order of magnitude. Although syndrome extraction involves nonlocal interactions, the required qubit routing can be implemented deterministically using global cyclic column shifts and only a few parallel row permutations. As a result, the routing time remains three orders of magnitude shorter than the coherence times already demonstrated in neutral-atom arrays.

Several important challenges remain. First, although the code automorphisms provide efficient primitives for logical-qubit routing, fault-tolerant quantum computation requires a universal set of logical operations. A promising route is lattice surgery, but constructing fault-tolerant surgery protocols for Cornucopia codes while respecting the rigid-motion constraints imposed by crossed AODs remains an open challenge.
Second, practical fault-tolerant quantum computing requires real-time decoding. Decoding large Cornucopia code blocks within the duration of a syndrome extraction cycle remains a notable challenge. Addressing this challenge will require faster decoding algorithms tailored to the specific structure of the Cornucopia codes, potentially leveraging modern AI techniques~\cite{Bausch2024Learning,Gu2026Scalable}. 
Third, it will be essential to develop end-to-end compilation pipelines and evaluate the full space-time cost of logical algorithms. Such resource estimates will clarify which quantum algorithms, logical gate constructions and compilation strategies can most effectively exploit the ultra-high rate and hardware-efficient structure of the Cornucopia codes.

\bibliography{qldpc}

\renewcommand{\figurename}{Extended Data Fig.}
\setcounter{figure}{0}
\renewcommand{\tablename}{Extended Data Tab.}
\setcounter{table}{0}

\clearpage
\noindent\textbf{\large{}Methods}

\noindent\textbf{Code construction} \\
Cornucopia codes constitute a highly structured subfamily of the qLDPC code constructions recently introduced in Ref.~\cite{Kasai2026Breaking} and subsequently adapted for reconfigurable atom arrays in Ref.~\cite{Zhao2026UltraHighRate}. 
In this work, we formalize Cornucopia codes through an explicit block-convolutional construction on a quasi-cyclic spatial geometry. 

The Cornucopia codes are defined over a two-dimensional discrete torus \(V=\mathbb Z_3\times\mathbb Z_q\), where the parameter \(q\) dictates the cycle length of the column registers.
The total number of coordinates per block is $P = 3q$, where $\gcd(q,3)=1$ ensures a valid Chinese Remainder Theorem decomposition of the underlying cyclic group.
As shown in Fig.~\ref{fig:1}\textbf{a}, the physical data qubits are partitioned into twelve disjoint blocks: six $L$-type blocks denoted $L_0, \dots, L_5$ and six $R$-type blocks denoted $R_0, \dots, R_5$. The auxiliary check qubits are correspondingly partitioned into six blocks: three $X$-type blocks $X_0, X_1, X_2$ and three $Z$-type blocks $Z_0, Z_1, Z_2$. Each block contains exactly $P$ qubits, yielding a total of $36q$ data qubits and $18q$ check qubits for a given code in the family.

The stabilizer code is fully specified by its $X$- and $Z$-type parity-check matrices, $H_X$ and $H_Z$, which we construct in a block-convolutional form. For check block indices $i \in \{0, 1, 2\}$ and data block indices $j \in \{0, 1, \dots, 5\}$, the sub-blocks of the parity-check matrices are given by
\begin{equation}
H_X = \big[ \mathbf{A} \mid \mathbf{B} \big], \qquad H_Z = \big[ \mathbf{\tilde{B}} \mid \mathbf{\tilde{A}} \big]
\end{equation}
where the individual $(i,j)$-th blocks explicitly read
\begin{equation}
\begin{aligned}
(\mathbf{A})_{i,j} &= A_{j-i}, & (\mathbf{B})_{i,j} &= B_{j-i}, \\
(\mathbf{\tilde{B}})_{i,j} &= B_{i-j}^{-1}, & (\mathbf{\tilde{A}})_{i,j} &= A_{i-j}^{-1}.
\end{aligned}
\end{equation}
Here, all block indices are evaluated modulo $6$.

The matrix elements $A_k$ and $B_k$ (for $k \in \mathbb{Z}_6$) are $P \times P$ permutation matrices, each representing a distinct coordinate permutation on the $\mathbb{Z}_3 \times \mathbb{Z}_q$ grid. Let a physical qubit within a block be indexed by its coordinate $(x,y)$. All arithmetic on the coordinate pairs $(x,y)$ is implicitly evaluated modulo $3$ for the row indices and modulo $q$ for the column indices.
The coordinate permutations are classified into three structural families corresponding to affine mappings on the row and column coordinates:
\begin{enumerate}
\item \textit{Row-inverting column translation:} The operators $A_1$ and $B_3$ implement an orientation-reversing transformation on the row index, defined by 
\begin{equation}
(x, y) \mapsto (2x + 2, y + s).
\label{eq:Row-inverting}
\end{equation}
This mapping swaps two of the three rows while cyclically shifting the columns. 
\item \textit{Row-column translation:} The operators $A_0$ and $B_2$ define a uniform translation in both dimensions, 
\begin{equation}
(x, y) \mapsto (x + 1, y + s),
\label{eq:Row-column}
\end{equation}
cyclically shifting both rows and columns. 
\item \textit{Pure column translation:} The remaining operators ($A_2, A_3, A_4, A_5$ and $B_0, B_1, B_4, B_5$) act as pure column shifts 
\begin{equation}
(x, y) \mapsto (x, y + s),
\label{eq:Pure column}
\end{equation}
leaving the row indices invariant.
\end{enumerate}
Across the three types of coordinate permutations, the column-shift parameters $\bm{s} \in \mathbb{Z}^{12}_q$ associated with each of the twelve operators are not constrained. They provide the search degrees of freedom and are optimized to improve code distance and logical performance.

By construction, the parity-check matrices satisfy the valid CSS code requirement $H_X H_Z^{\top} = 0 \pmod 2$. 
Evaluating the inner product between the $i$-th block-row of $H_X$ and the $k$-th block-row of $H_Z$ reduces this requirement to the explicit algebraic constraint
\begin{equation}
\sum_{j=0}^5 \left( A_{j-i} B_{k-j} + B_{j-i} A_{k-j} \right) = 0 \pmod 2
\label{eq:css}
\end{equation}
for all $i, k \in \{0, 1, 2\}$. 
In our geometrical layout, because the majority of the coordinate permutations are pure translations, they pairwise commute.
The only residual non-commuting operator pairs are $(A_1, B_2)$ and $(A_0, B_3)$, which stem from the interplay between the row-inverting column translations and the row-column translations. However, these non-commuting pairs never appear in the summand term of Eq.~\ref{eq:css}. Thus, all terms participating in the summation strictly commute and pairwise cancel under modulo-$2$ arithmetic.

Because the $12P$ physical data qubits are constrained by at most $6P$ auxiliary checks, the encoding rate of the Cornucopia code family is bounded below by $1/2$. Any linear dependencies within the stabilizer group will contribute additional logical degrees of freedom, thereby further elevating the exact code rate.

\vspace{2mm}
\noindent\textbf{Exact code distance certification} \\
Let $\operatorname{wt}(v)$ denote the Hamming weight, $\ker(M)$ denote the null space comprising all binary vectors $v$ such that $Mv = 0 \pmod 2$, and $\operatorname{rs}(M)$ denote the row space spanned by the rows of a matrix $M$.
\begin{lemma} A Cornucopia code offers exactly equal distance for $X$-type and $Z$-type errors, $d_X = d_Z = d$, where the respective distances are defined as
\begin{equation*}
\begin{aligned}
d_X &= \min\{\operatorname{wt}(v) : v \in \ker(H_Z) \setminus \operatorname{rs}(H_X)\}, \\
d_Z &= \min\{\operatorname{wt}(v) : v \in \ker(H_X) \setminus \operatorname{rs}(H_Z)\}.
\end{aligned}
\end{equation*}
\end{lemma}
The proof is provided in the Supplementary Information. As $d_X=d_Z$, we rigorously certify the exact code distances by evaluating the minimal weights of non-trivial logical operators only in the $X$ basis. 

A distance claim is accepted only after it is exhaustively certified: all candidate operators of weight strictly less than $d$ must be completely excluded in both Pauli bases, and an explicit logical witness at weight $d$ must be found that commutes with all opposite-type checks while lying strictly outside the corresponding stabilizer row space.
Notably, the code distance is strictly an even integer. Because summing all rows of $H_Z$ yields an all-ones vector, the all-ones operator is inherently a stabilizer constraint. Thus, any valid logical operator must commute with this all-ones vector, which strictly requires its Hamming weight to be even.

To circumvent the combinatorial explosion inherent in brute-force distance verification, we implement an exact, symmetry-reduced search algorithm based on Tanner graph row-pairing enumeration. Within our specific block architecture (comprising $12$ data blocks and $3$ check blocks per basis), each physical data qubit is incident to exactly one check node within any check block. Thus, any valid zero-syndrome logical operator must intersect each check node an even number of times, allowing us to decompose its support strictly into same-row qubit pairs. Since each check node is connected to exactly $12$ data blocks, it permits $\binom{12}{2} = 66$ valid intra-check pairing combinations. This bipartite structural property drastically reduces the search space: evaluating a weight-$2t$ logical operator no longer requires searching through an arbitrary $\binom{12P}{2t}$ physical supports, but rather scales down to selecting $t$ pairs from a total of $66P$ legal pair configurations.

Furthermore, we exploit the geometrical coordinate automorphisms of the Cornucopia codes to compact the symmetry orbits. By quotienting out the $q$-fold cyclic column translations, we reduce the number of inequivalent root qubit pairs, which define the starting branches of our search tree, from $66P=198q$ down to $66\times3=198$. By locking the first pair within this restricted orbit representative, the search space upper bound for a weight-$2t$ operator is structurally compressed to $198 \times \binom{66P-1}{t-1}$.

During this complete enumeration, we also perform dynamic residual-syndrome pruning and stabilizer row-space reductions. In particular, we prune any search path whose residual syndrome cannot be eliminated by the remaining weight allowance. For example, because each physical data qubit is incident to exactly three check nodes (one within each of the three check blocks per basis), adding $\Delta w$ further qubits to the search path can alter at most $3\Delta w$ syndrome bits. If a partial configuration leaves a residual syndrome with strictly more than $3\Delta w$ unsatisfied checks, it becomes mathematically impossible to resolve them and return to a global zero-syndrome state, prompting the immediate abortion of that branch. 
In addition, we terminate branches that exhibit coset equivalence to known stabilizers. Together, these algorithmic accelerations conclusively guarantee the exact code distance without omitting any possible non-trivial logical operators.

\vspace{2mm}
\noindent\textbf{Syndrome extraction schedule} \\
The syndrome extraction schedule (shown in Fig.~\ref{fig:1}\textbf{c}) is explicitly designed to preserve stabilizer determinism. This property can be formalized through the following proposition.
\begin{proposition} 
For any Cornucopia code, the following syndrome extraction schedule, comprising twelve layers of non-overlapping CNOT gates:
\begin{equation*}
\begingroup
\setlength{\arraycolsep}{1pt}
\begin{array}{r|cccccc|cccccc}
\hline
t: 
& 0 & 1 & 2 & 3 & 4 & 5 
& 6 & 7 & 8 & 9 & 10 & 11 \\
\hline
X: 
& A_0 & A_1 & A_2 & A_3 & A_4 & A_5 
& B_4 & B_5 & B_0 & B_1 & B_2 & B_3 \\
Z: 
& B_3^{-1} & B_2^{-1} & B_1^{-1} & B_0^{-1} & B_5^{-1} & B_4^{-1} 
& A_5^{-1} & A_4^{-1} & A_3^{-1} & A_2^{-1} & A_1^{-1} & A_0^{-1} \\
\hline
\end{array}
\endgroup
\end{equation*}
avoids all intra-layer data block conflicts. Furthermore, the parity of the relative time ordering (e.g., $X$-before-$Z$) across all overlaps between any $X$-type and $Z$-type check is always even, guaranteeing deterministic stabilizer readouts.
\end{proposition}

The physical intuition underlying this proposition relies on the precise block-cyclic topology of the Cornucopia code. First, the absence of data block conflicts is guaranteed, as the schedule offsets the targeted data blocks for the $X$- and $Z$-type checks by exactly three blocks modulo $6$. Thus, the two check types are always routed to disjoint halves of the $L$- or $R$-type data blocks, physically preventing any simultaneous CNOT operations on the same data qubit. 
Second, whenever an $X$-check and a $Z$-check share support across the overall schedule, the code's bilateral layout dictates that their overlaps always emerge in exact pairs---one within the left data blocks and its twin within the right data blocks. With all overlapping checks structurally guaranteed to commute, the temporal schedule enforces an identical relative CNOT ordering ($X$-before-$Z$ or $Z$-before-$X$) for both overlaps in a pair. Thus, their measurement parity contributions perfectly cancel out under modulo-$2$ arithmetic. The rigorous mathematical proof is detailed in the Supplementary Information.

\vspace{2mm}
\noindent\textbf{Circuit-level noise simulation and decoding} \\
We evaluate the logical performance of Cornucopia codes by simulating logical memory experiments. Each experiment consists of $N_{c}$ repeated syndrome extraction cycles, concluding with a transversal destructive measurement of all data qubits in the chosen logical basis. 
We employ a standard circuit-level depolarizing noise model, which assumes that independent Pauli errors are applied to each component operation within the syndrome measurement circuit. 
We parameterize the model using a uniform physical error rate, $p$, applied across all operations, including initial data qubit preparation, CNOT gates, measurements of check qubits, reset of check qubits, the final readout of data qubits.
Specifically, a faulty CNOT gate is modeled by applying an ideal CNOT gate followed by a two-qubit Pauli error drawn uniformly from the $15$ non-identity elements of $\{I, X, Y, Z\}^{\otimes 2}$, each occurring with probability $p/15$. State preparation and reset faults are modeled by applying a bit-flip ($X$) error immediately after initialization in the $\lvert 0\rangle$ state, or a phase-flip ($Z$) error after initialization in the $\lvert +\rangle$ state, each with probability $p$. 
Similarly, measurement faults are represented by applying a basis-flip error immediately prior to an ideal measurement: an $X$ error before a $Z$-basis measurement, and a $Z$ error before an $X$-basis measurement, each occurring with probability $p$.

Given an error syndrome $\mathbf{s}$, we employ a hierarchical, three-stage decoding strategy. To maximize computational throughput, the primary relay-BP phase is partitioned into two sequential passes. We initially execute relay-BP with a small iteration cap to rapidly resolve the most typical error syndromes. 
Let $H_{\mathrm{dec}}$ denote the corresponding circuit-level parity-check matrix; a decoding pass is deemed to converge to an error estimate $\hat{\mathbf{e}}$ if it satisfies the syndrome condition 
\begin{equation*}
H_{\mathrm{dec}}\hat{\mathbf{e}} = \mathbf{s} \pmod 2.
\end{equation*}
If this rapid initial pass fails to converge, we trigger a secondary relay-BP pass with a larger iteration limit to tackle more stubborn error configurations. Finally, if the extended relay-BP pass still fails to find a valid error estimate $\hat{\mathbf{e}}$, we invoke BP-OSD as a fallback decoder.
This cascaded decoding strategy optimizes simulation efficiency by resolving the majority of samples with minimal iterations, reserving heavy OSD post-processing exclusively for challenging error configurations.

\vspace{2mm}
\noindent\textbf{Numerical simulation details} \\
It is conventional in fault-tolerant simulations to evaluate a distance-$d$ code over $N_c = d$ syndrome cycles. However, due to the large block lengths of Cornucopia codes, maintaining $N_c = d$ generates a massive space-time decoding graph, rendering offline decoding computationally prohibitive. To optimize simulation efficiency, we perform truncated cycle simulations. Specifically, we set $N_c = 6$ for all codes with $d \leq 12$, $N_c = 7$ for the $d = 14$ instance, and $N_c = 8$ for the largest code instances with $d \geq 16$. For consistency, this truncated cycle simulation is also applied to surface codes and bivariate bicycle codes.

We fit the logical error rate data presented in Fig.~\ref{fig:2} using the formula $p_L = p^{d/2}\exp(c_0 + c_1p + c_2p^2)$. This model is adopted for evaluating the sub-threshold scaling of surface and bivariate bicycle codes~\cite{Bravyi2013Simulation,Bravyi2024Highthreshold}. The extracted fitting parameters $c_i$ for all considered codes are provided in the Supplementary Information.

The circuit-level distance is defined as the minimum number of physical faults within the syndrome measurement circuit required to trigger an undetectable logical error. Following the methodology introduced in ref.~\cite{Bravyi2024Highthreshold}, we employ BP-OSD to compute upper bounds on the circuit-level distances for the Cornucopia code family. The resulting upper bounds are summarized in Tab.~\ref{tab:family}.

\vspace{2mm}
\noindent\textbf{Implementation in neutral-atom arrays} \\
To implement a Cornucopia code, each block is physically arranged as a $3\times q$ grid, with the twelve data blocks and the six check blocks stacked vertically to form the global 2D array, as illustrated in Extended Data Fig.~\ref{fig:atom_moving}. 
During each syndrome extraction layer, both the data and check arrays undergo targeted physical reconfiguration. Specifically, the data blocks are cyclically shifted as intact units to align with their designated check blocks, whereas the check blocks undergo internal coordinate permutations. This synchronized choreography brings all interacting qubit pairs into close gate-range proximity, enabling a parallel coordinate-wise transversal CNOT layer across the entire array.

For the syndrome extraction schedule (Fig.~\ref{fig:1}\textbf{c}), transitioning from layer $t$ to $t+1$ requires the check blocks to undergo a physical rearrangement governed by the transformation $M_{t+1}^{-1} M_t$, where $M_t$ denotes the coordinate permutation mapping between the check and data blocks in layer $t$. By design, all three types of coordinate permutations in Cornucopia codes strictly manifest as cyclic shifts in the column ($y$) dimension, limiting the necessary column routing to simple bulk cyclic shifts. 
In the row ($x$) dimension, eight of the permutations act as the identity, while the remaining four induce row shifts or swaps. 
Thus, row permutations are only required in specific syndrome layers and involve rearranging at most three rows, incurring relatively low routing overhead. The entire syndrome extraction cycle thus compiles down to a deterministic, repetitive sequence of bulk cyclic shifts and minor row permutations. 
This structural property is inherently compatible with reconfigurable neutral-atom arrays driven by crossed acousto-optic deflectors (AODs), whose separable control architecture enables concurrent translations along rows and columns~\cite{Xu2024Constantoverhead,Zhao2026UltraHighRate}.

The total duration of a full syndrome extraction cycle comprises three components: the execution of $12$ entangling layers, the readout of the check qubits, and the physical atom rearrangement between consecutive layers. In standard neutral-atom setups, a transversal CNOT gate is realized by a CZ gate interleaved with single-qubit Hadamard pulses, taking approximately $1\mu$s. This gate duration is fundamentally negligible compared to the atom transport time, which spans hundreds of microseconds. The measurement of the check qubits, which takes roughly $500\mu$s, can be pipelined with the reloading of fresh check qubits for the subsequent cycle, effectively hiding its time overhead. Thus, the total cycle time is dominated by the atom routing overhead between consecutive entangling layers.

We categorize the required atom routing into four types of movement: 
\begin{enumerate}
\item bulk cyclic shifts in the column direction within the check blocks;
\item bulk cyclic shifts in the row direction or row swapping within the check blocks;
\item vertical cyclic shifts of the data blocks;
\item macroscopic swapping of all $L$-type and $R$-type data blocks. 
\end{enumerate}
In our conservative scheduling model, we allocate two crossed-AODs to each cyclic shift or macroscopic swap, ensuring that the bulk motion and the boundary wrap-around, or the opposing trajectories of a swap, can be executed concurrently without inducing atom collisions. 
Assuming four crossed-AODs are available, a maximum of two independent atom routing operations can be executed concurrently.  As the cyclic shifts for the $X$- and $Z$-type check blocks generally involve different displacement distances, executing them simultaneously fully occupies the four crossed-AODs. To adhere to this hardware concurrency limit, we sequence the physical reconfiguration within each transition step into distinct, highly parallelized batches. First, we execute the data-block routing, which comprises the vertical cyclic shifts of the data blocks and the macroscopic swapping of all $L$- and $R$-type blocks, the latter occurring exclusively in transition steps $(5)$ and $(11)$. Subsequently, as illustrated in Extended Data Fig.~\ref{fig:atom_moving}, we execute the row permutations for both the $X$- and $Z$-type check blocks in parallel; these operations are only required during transition steps $(0)$, $(1)$, $(9)$, and $(10)$. Finally, we concurrently perform the bulk cyclic shifts in the column direction for both the $X$- and $Z$-type check blocks.
Thus, the total transport time $\mathcal{T}_i$ during transition step $i$ can be estimated as:
\begin{equation}
\begin{aligned}
\mathcal{T}_i =& T^{i}_{\text{data,shift}} + T^{i}_{\text{data,swap}} + T^{i}_{\text{row}} + T^{i}_{\text{col}} \\
=& T^{i}_{\text{data,shift}} + T^{i}_{\text{data,swap}} 
+ \max\left\{ T^{i}_{X,\text{row}}, T^{i}_{Z,\text{row}} \right\} \\
&+
\max\left\{T^{i}_{X,\text{col}}, T^{i}_{Z,\text{col}}
\right\},
\end{aligned}
\label{eq:estimate}
\end{equation}
where $T^{i}_{\text{data,shift}}$ and $T^{i}_{\text{data,swap}}$ represent the time required for the vertical cyclic shifts and the macroscopic swapping of the data blocks, respectively. The terms $T^{i}_{X,\text{row}}$ and $T^{i}_{Z,\text{row}}$ denote the durations of the row permutations for the $X$- and $Z$-type check blocks, while $T^{i}_{X,\text{col}}$ and $T^{i}_{Z,\text{col}}$ correspond to the time costs of their respective bulk cyclic shifts in the column direction. For any routing operation not required in a given step $i$, its corresponding time overhead is exactly zero. The total atom routing time overhead per syndrome extraction cycle is subsequently obtained by summing $\mathcal{T}_i$ over all transition steps.

To evaluate these discrete components, we employ a constant-acceleration and deceleration kinematic model ~\cite{Xu2024Constantoverhead,Zhao2026UltraHighRate}. 
The pure flight time for a linear displacement $D$ is given by $t_{\text{fly}}(D) = 2\sqrt{D/a_{\text{max}}}$, where $a_{\max}$ is the peak atom moving acceleration. Accounting for the trap transfer time $\tau_t$ required to transfer atoms between the static and dynamic optical tweezers, a standard block displacement takes $2\tau_t + t_{\text{fly}}(D)$. 
Crucially, for cyclic shifts within a chain of length $C$ that execute a $\Delta s>0$ bulk shift alongside an $(C-\Delta s)$ boundary wrap-around, the concurrent execution using independent AODs takes
\begin{equation}
2\tau_t +\max \left\{t_{\text{fly}}(\Delta s \cdot d_0), t_{\text{fly}}((C-\Delta s) \cdot d_0) \right\}, 
\label{eq:shift}
\end{equation}
where $d_0$ is the uniform atom array lattice spacing. 

To calculate each term in Eq.~\ref{eq:estimate}, we utilize parameters demonstrated in recent experiments~\cite{Bluvstein2022Quantum}: $d_0 = 12\ \mu\text{m}$, $a_{\max}=0.0055\ \mu\text{m}/\mu\text{s}^2$, and $\tau_t=50\ \mu\text{s}$. 
The term $T^{i}_{\text{data,shift}}$ is required in all transition steps except for step $(5)$. Since the vertical cyclic shifts of the data blocks traverse a fixed span of $15$ rows ($180\mu\text{m}$), this time overhead is a constant. Using Eq.~\ref{eq:shift}, it is evaluated as $2\tau_t + t_{\text{fly}}(180\mu\text{m}) \approx 461.8\mu\text{s}$. 
The macroscopic block swapping time, $T^{i}_{\text{data,swap}}$, occurs exclusively in steps $(5)$ and $(11)$. It is similarly a constant, moving across $18$ rows ($216\mu\text{m}$), which evaluates to $2\tau_t + t_{\text{fly}}(216\ \mu\text{m}) \approx 496.3\mu\text{s}$. Aggregating the $11$ vertical cyclic shifts and the $2$ macroscopic data block swaps across a full syndrome cycle yields a total routing time for data blocks of approximately $6.07\text{ms}$. Note that this portion of the overhead is an architectural constant for all Cornucopia codes, independent of the code size.

In contrast, the ancilla routing term in Eq.~\ref{eq:estimate} is intrinsically dependent on the specific Cornucopia code instance. The required column-shift displacements are dictated by the shift parameter $\bm{s} \in \mathbb{Z}^{12}_q$, as defined in Eqs.~\ref{eq:Row-inverting}, \ref{eq:Row-column}, and \ref{eq:Pure column}. Since the maximum wrap-around distance grows linearly with the grid dimension $q$, the corresponding kinematic routing overhead inherently scales with the code size as $O(\sqrt{q})$. By evaluating this dynamic term for each code instance presented in Tab.~\ref{tab:family}, we determine the total duration of a full syndrome extraction cycle across the Cornucopia code family. These time estimates are summarized in Extended Tab.~\ref{tab:movement_cost}.

\vspace{2mm}
\noindent\textbf{Logical routing} \\
Efficient logical routing, the transport of logical observables to interaction zones or lattice surgery boundaries, is essential for scalable fault-tolerant quantum computation. 
Cornucopia codes provide a native, low-overhead routing mechanism by exploiting their underlying quasi-cyclic symmetry, avoiding deep routing circuits or resource-intensive code deformation protocols. 
As visualized in Fig.~\ref{fig:3}, the column-shift automorphism preserves the stabilizer group and acts as a permutation on the encoded logical space.  
We formalize the algebraic structure of this native logical routing mechanism in the following proposition. The proof is provided in the Supplementary Information.

To formally characterize the logical action, let $T_c \colon (b,r,c) \mapsto (b,r,c+1)$ denote the global column-shift permutation, where $b$ is the data-block label, $r\in \mathbb Z_3$ is the row coordinate, and $c\in \mathbb Z_q$ is the column coordinate. We define the $X$- and $Z$-type logical Pauli quotient spaces as
\begin{equation*}
\mathcal L_X=\ker H_Z/\operatorname{row}(H_X),
\qquad
\mathcal L_Z=\ker H_X/\operatorname{row}(H_Z).
\end{equation*}
Since $T_c$ preserves the stabilizer group, it induces linear automorphisms $A_X\colon \mathcal L_X\to\mathcal L_X$ and $A_Z\colon \mathcal L_Z\to\mathcal L_Z$. The property $T_c^q=I$ naturally implies $A_X^q=I$ and $A_Z^q=I$. Letting $C_q=\langle u\mid u^q=1\rangle$ be the cyclic group generated by the column-shift action, we equip $\mathcal L_X$ and $\mathcal L_Z$ with $\mathbb F_2[C_q]$-module structures by defining the action of the generator $u$ as $A_X$ on $\mathcal L_X$ and as $A_Z$ on $\mathcal L_Z$. 
Equivalently, the group algebra is isomorphic to the polynomial quotient ring $\mathbb F_2[C_q]\simeq \mathbb F_2[u]/(u^q-1)$. As a module over itself, $\mathbb F_2[C_q]$ is the regular module of dimension $q$ over $\mathbb F_2$, where $u$ acts as a length-$q$ cyclic shift. 
Throughout, let $\mathbf{1}$ denote the one-dimensional trivial module (on which $u$ acts as the identity), and let $m\cdot M$ denote the direct sum of $m$ copies of a module $M$.

\begin{proposition}[Logical action of the column-shift automorphism]
For the Cornucopia code instances in Tab.~\ref{tab:family} of the main text, all of which satisfy $k=n/2+4$, the induced action of $T_c$ on both logical spaces yields the module decomposition
\begin{equation}
\mathcal L_{X/Z}\simeq
18\cdot \mathbb F_2[C_q]\oplus 4\cdot \mathbf{1}.
\end{equation}
Equivalently, there exist symmetry-adapted bases for $\mathcal{L}_X$ and $\mathcal{L}_Z$ individually, such that within each space, a single physical column shift implements eighteen simultaneous logical cyclic shifts of length $q$, together with four fixed logical modes.
\end{proposition}

\vspace{.4cm}
\noindent\textbf{\large{}Data availability}\\
The data presented in the figures and that support the other findings of this study will be made publicly available for download on Zenodo upon publication.

\vspace{.4cm}
\noindent\textbf{\large{}Code availability}\\
The data analysis and numerical simulation codes will be made publicly available for download on Zenodo upon publication.

\vspace{.4cm}
\noindent\textbf{\large{}Acknowledgments}\\
The authors thank Zheng-Zhi Sun, Yinghao Ma, Hanyu Yan, Bingrun Wang, and Boren Gu for helpful discussions. 
This work is supported by the National Natural Science Foundation of China (T2225008), the Quantum Science
and Technology-National Science and Technology Major Project (2021ZD0302203), the Fundamental and Interdisciplinary Disciplines Breakthrough Plan of the Ministry of Education of China (JYB2025XDXM112), the Shanghai Qi Zhi Institute Innovation Program (SQZ202318), and the Tsinghua University Dushi Program. 

\vspace{.4cm}
\noindent\textbf{\large{}Author contributions} \\
Z.L. and W.L. designed quantum error correction codes, the syndrome measurement circuit, and decoding algorithms used in this study. 
Z.L. and W.L. designed atom rearrangement protocol and estimated atom-routing time for repeated syndrome extraction under the supervision of D.-L.D..
Z.L. performed the numerical simulations. All authors contributed to writing the manuscript.

\vspace{.4cm}
\noindent\textbf{\large{}Competing interests} \\
All authors declare no competing interests.

\begin{table*}[t]
\label{tab:code_parameters}
\centering
\caption{\textbf{Parameters of Cornucopia codes.} All listed codes have weight-$12$ checks and a depth-$12$ syndrome measurement circuit. The tuple $[[n,k,d]]$ denotes the number of physical data qubits, logical qubits, and the code distance, respectively. The parameter $q=n/36$ represents the column dimension of the atomic grids for each check or data block. The coordinate permutations $A_i$ and $B_i$ manifest as cyclic shifts in the column ($y$) dimension. The specific shift amounts $s$ modulo $q$, denoted as $(~~, y+s)$ in the table, are formally defined in Eqs.~\ref{eq:Row-inverting}, \ref{eq:Row-column}, and \ref{eq:Pure column}.}
\renewcommand{\arraystretch}{1}
\setlength{\tabcolsep}{16pt}
\begin{tabular}{c c c c c}
\toprule
{\bfseries\boldmath$[[n,k,d]]$}
& {\bfseries\boldmath$q$}
& $i$ 
& $A_{i}:(~~,y+s)$ 
& $B_i:(~~,y+s)$ \\
\midrule

\multirow{6}{*}{$[[252,130,6]]$}
& \multirow{6}{*}{$7$}
& $0$ & $y+2$  & $y+5$ \\
& & $1$ & $y+1$  & $y+3$ \\
& & $2$ & $y+1$  & $y$ \\
& & $3$ & $y+1$  & $y+5$ \\
& & $4$ & $y+4$  & $y+2$ \\
& & $5$ & $y+5$  & $y+3$ \\
\midrule

\multirow{6}{*}{$[[576,292,8]]$}
& \multirow{6}{*}{$16$}
& $0$ & $y+1$  & $y+6$ \\
& & $1$ & $y+5$  & $y+12$ \\
& & $2$ & $y+10$ & $y+10$ \\
& & $3$ & $y+12$ & $y+11$ \\
& & $4$ & $y+12$ & $y+6$ \\
& & $5$ & $y+8$  & $y+3$ \\
\midrule

\multirow{6}{*}{$[[900,454,10]]$}
& \multirow{6}{*}{$25$}
& $0$ & $y+23$ & $y+9$ \\
& & $1$ & $y+7$  & $y+7$ \\
& & $2$ & $y+7$  & $y+11$ \\
& & $3$ & $y+12$ & $y+17$ \\
& & $4$ & $y+19$ & $y+10$ \\
& & $5$ & $y+6$  & $y+7$ \\
\midrule

\multirow{6}{*}{$[[1044,526,12]]$}
& \multirow{6}{*}{$29$}
& $0$ & $y+2$  & $y+27$ \\
& & $1$ & $y+22$ & $y+11$ \\
& & $2$ & $y+20$ & $y+12$ \\
& & $3$ & $y+22$ & $y+18$ \\
& & $4$ & $y+18$ & $y+21$ \\
& & $5$ & $y+6$  & $y+26$ \\
\midrule

\multirow{6}{*}{$[[1764,886,14]]$}
& \multirow{6}{*}{$49$}
& $0$ & $y+44$ & $y+42$ \\
& & $1$ & $y+5$  & $y+38$ \\
& & $2$ & $y+33$ & $y+32$ \\
& & $3$ & $y+25$ & $y+46$ \\
& & $4$ & $y+22$ & $y+45$ \\
& & $5$ & $y+7$  & $y+29$ \\
\midrule

\multirow{6}{*}{$[[2304,1156,16]]$}
& \multirow{6}{*}{$64$}
& $0$ & $y+19$ & $y+20$ \\
& & $1$ & $y+54$ & $y+21$ \\
& & $2$ & $y+25$ & $y+48$ \\
& & $3$ & $y+4$  & $y+14$ \\
& & $4$ & $y+7$  & $y+56$ \\
& & $5$ & $y+51$ & $y+55$ \\
\midrule

\multirow{6}{*}{$[[2844,1426,18]]$}
& \multirow{6}{*}{$79$}
& $0$ & $y+6$  & $y+24$ \\
& & $1$ & $y+49$ & $y+41$ \\
& & $2$ & $y+55$ & $y+78$ \\
& & $3$ & $y+18$ & $y+53$ \\
& & $4$ & $y+40$ & $y+68$ \\
& & $5$ & $y+7$  & $y+21$ \\
\bottomrule
\end{tabular}
\end{table*}

\begin{table*}[t]
\centering
\caption{\textbf{Estimates of atom-routing time per syndrome cycle for Cornucopia codes.} 
The required ancilla atom routing time (in $\mu\text{s}$) is detailed for each transition step in Extended Data Figs.~\ref{fig:atom_moving} and \ref{fig:atom_moving2}. For transition steps that require both row permutations and bulk cyclic shifts in the column direction (e.g., steps $0, 1, 9, \text{and } 10$), the overhead is presented as the sum of their sequential execution durations ($T^{i}_{\text{row}} + T^{i}_{\text{col}}$). Owing to the structural symmetry of the scheduling in Fig.~\ref{fig:1}\textbf{c}, the step-wise ancilla overheads exhibit a temporal reflection symmetry between step $(i)$ and step $(10-i)$. 
Step $(11)$ requires no atom routing for check blocks.
The total data-block routing overhead remains an architectural constant of $6.07\ \text{ms}$, independent of the code size. The final cycle time represents the aggregate duration of both the ancilla and data routing operations required to complete one full syndrome cycle.
}
\label{tab:movement_cost}
\renewcommand{\arraystretch}{1.5}
\setlength{\tabcolsep}{4pt}
\resizebox{\textwidth}{!}{
\begin{tabular}{c c c c c c c c}
\toprule
Step ($i$) 
& $[[252, 130, 6]]$ 
& $[[576, 292, 8]]$ 
& $[[900, 454, 10]]$ 
& $[[1044, 526, 12]]$ 
& $[[1764, 886, 14]]$ 
& $[[2304, 1156, 16]]$ 
& $[[2844, 1426, 18]]$ \\
\midrule
$0, 10$ & $232.1 + 328.8$ & $232.1 + 461.8$ & $232.1 + 507.2$ & $232.1 + 548.0$ & $232.1 + 683.4$ & $232.1 + 652.7$ & $232.1 + 786.5$ \\
$1, 9$  & $232.1 + 286.8$ & $232.1 + 449.5$ & $232.1 + 528.1$ & $232.1 + 594.3$ & $232.1 + 712.6$ & $232.1 + 668.2$ & $232.1 + 898.2$ \\
$2, 8$  & $308.9$ & $449.5$ & $548.0$ & $585.4$ & $726.7$ & $841.5$ & $835.6$ \\
$3, 7$  & $308.9$ & $436.8$ & $548.0$ & $594.3$ & $733.6$ & $829.6$ & $914.4$ \\
$4, 6$  & $328.8$ & $436.8$ & $538.2$ & $557.7$ & $644.7$ & $841.5$ & $740.5$ \\
$5$     & $286.8$ & $449.5$ & $528.1$ & $461.8$ & $675.9$ & $817.6$ & $829.6$ \\
\midrule
Ancilla Total (ms) & $4.34$ & $5.85$ & $6.80$ & $7.15$ & $8.61$ & $9.41$ & $10.11$ \\
Data Total (ms) & $6.07$ & $6.07$ & $6.07$ & $6.07$ & $6.07$ & $6.07$ & $6.07$ \\
\midrule
\textbf{Cycle Total (ms)} & \textbf{10.41} & \textbf{11.92} & \textbf{12.87} & \textbf{13.22} & \textbf{14.68} & \textbf{15.48} & \textbf{16.18} \\
\bottomrule
\end{tabular}
}
\end{table*}

\begin{table*}[t]
\centering
\caption{
\textbf{Syndrome extraction schedule for Cornucopia codes.} 
A full syndrome measurement cycle contains twelve sequential layers of transversal CNOT gates, indexed from $0$ to $11$. ``Trans-CNOT'' denotes the transversal CNOT gates applied between a specified check block and a data block, with the exact physical qubit coordinate pairings dictated by $A_i, B_i$ and $A_i^{-1}, B_i^{-1}$.
}
\label{tab:depth12_syndrome_schedule}
\renewcommand{\arraystretch}{1.5}
\setlength{\tabcolsep}{12pt}

\begin{tabular}{c l c l c l}
\toprule
{\bfseries Layer} & {\bfseries Circuit}
& {\bfseries Layer} & {\bfseries Circuit} 
& {\bfseries Layer} & {\bfseries Circuit} \\
\midrule

$0$ &
\begin{tabular}[t]{@{}l@{}}
Trans-CNOT: \\
\quad $X_0 \rightarrow L_0$ \quad $[A_0]$\\
\quad $X_1 \rightarrow L_1$ \quad $[A_0]$\\
\quad $X_2 \rightarrow L_2$ \quad $[A_0]$\\
\quad $L_3 \rightarrow Z_0$ \quad $[B_3^{-1}]$\\
\quad $L_4 \rightarrow Z_1$ \quad $[B_3^{-1}]$\\
\quad $L_5 \rightarrow Z_2$ \quad $[B_3^{-1}]$
\end{tabular}
&
$1$ &
\begin{tabular}[t]{@{}l@{}}
Trans-CNOT: \\
\quad $X_0 \rightarrow L_1$ \quad $[A_1]$\\
\quad $X_1 \rightarrow L_2$ \quad $[A_1]$\\
\quad $X_2 \rightarrow L_3$ \quad $[A_1]$\\
\quad $L_4 \rightarrow Z_0$ \quad $[B_2^{-1}]$\\
\quad $L_5 \rightarrow Z_1$ \quad $[B_2^{-1}]$\\
\quad $L_0 \rightarrow Z_2$ \quad $[B_2^{-1}]$
\end{tabular}
&
$2$ &
\begin{tabular}[t]{@{}l@{}}
Trans-CNOT: \\
\quad $X_0 \rightarrow L_2$ \quad $[A_2]$\\
\quad $X_1 \rightarrow L_3$ \quad $[A_2]$\\
\quad $X_2 \rightarrow L_4$ \quad $[A_2]$\\
\quad $L_5 \rightarrow Z_0$ \quad $[B_1^{-1}]$\\
\quad $L_0 \rightarrow Z_1$ \quad $[B_1^{-1}]$\\
\quad $L_1 \rightarrow Z_2$ \quad $[B_1^{-1}]$
\end{tabular}
\\
\midrule

$3$ &
\begin{tabular}[t]{@{}l@{}}
Trans-CNOT: \\
\quad $X_0 \rightarrow L_3$ \quad $[A_3]$\\
\quad $X_1 \rightarrow L_4$ \quad $[A_3]$\\
\quad $X_2 \rightarrow L_5$ \quad $[A_3]$\\
\quad $L_0 \rightarrow Z_0$ \quad $[B_0^{-1}]$\\
\quad $L_1 \rightarrow Z_1$ \quad $[B_0^{-1}]$\\
\quad $L_2 \rightarrow Z_2$ \quad $[B_0^{-1}]$
\end{tabular}
&
$4$ &
\begin{tabular}[t]{@{}l@{}}
Trans-CNOT: \\
\quad $X_0 \rightarrow L_4$ \quad $[A_4]$\\
\quad $X_1 \rightarrow L_5$ \quad $[A_4]$\\
\quad $X_2 \rightarrow L_0$ \quad $[A_4]$\\
\quad $L_1 \rightarrow Z_0$ \quad $[B_5^{-1}]$\\
\quad $L_2 \rightarrow Z_1$ \quad $[B_5^{-1}]$\\
\quad $L_3 \rightarrow Z_2$ \quad $[B_5^{-1}]$
\end{tabular}
&
$5$ &
\begin{tabular}[t]{@{}l@{}}
Trans-CNOT: \\
\quad $X_0 \rightarrow L_5$ \quad $[A_5]$\\
\quad $X_1 \rightarrow L_0$ \quad $[A_5]$\\
\quad $X_2 \rightarrow L_1$ \quad $[A_5]$\\
\quad $L_2 \rightarrow Z_0$ \quad $[B_4^{-1}]$\\
\quad $L_3 \rightarrow Z_1$ \quad $[B_4^{-1}]$\\
\quad $L_4 \rightarrow Z_2$ \quad $[B_4^{-1}]$
\end{tabular}
\\
\midrule

$6$ &
\begin{tabular}[t]{@{}l@{}}
Trans-CNOT: \\
\quad $X_0 \rightarrow R_4$ \quad $[B_4]$\\
\quad $X_1 \rightarrow R_5$ \quad $[B_4]$\\
\quad $X_2 \rightarrow R_0$ \quad $[B_4]$\\
\quad $R_1 \rightarrow Z_0$ \quad $[A_5^{-1}]$\\
\quad $R_2 \rightarrow Z_1$ \quad $[A_5^{-1}]$\\
\quad $R_3 \rightarrow Z_2$ \quad $[A_5^{-1}]$
\end{tabular}
&
$7$ &
\begin{tabular}[t]{@{}l@{}}
Trans-CNOT: \\
\quad $X_0 \rightarrow R_5$ \quad $[B_5]$\\
\quad $X_1 \rightarrow R_0$ \quad $[B_5]$\\
\quad $X_2 \rightarrow R_1$ \quad $[B_5]$\\
\quad $R_2 \rightarrow Z_0$ \quad $[A_4^{-1}]$\\
\quad $R_3 \rightarrow Z_1$ \quad $[A_4^{-1}]$\\
\quad $R_4 \rightarrow Z_2$ \quad $[A_4^{-1}]$
\end{tabular}
&
$8$ &
\begin{tabular}[t]{@{}l@{}}
Trans-CNOT: \\
\quad $X_0 \rightarrow R_0$ \quad $[B_0]$\\
\quad $X_1 \rightarrow R_1$ \quad $[B_0]$\\
\quad $X_2 \rightarrow R_2$ \quad $[B_0]$\\
\quad $R_3 \rightarrow Z_0$ \quad $[A_3^{-1}]$\\
\quad $R_4 \rightarrow Z_1$ \quad $[A_3^{-1}]$\\
\quad $R_5 \rightarrow Z_2$ \quad $[A_3^{-1}]$
\end{tabular}
\\
\midrule

$9$ &
\begin{tabular}[t]{@{}l@{}}
Trans-CNOT: \\
\quad $X_0 \rightarrow R_1$ \quad $[B_1]$\\
\quad $X_1 \rightarrow R_2$ \quad $[B_1]$\\
\quad $X_2 \rightarrow R_3$ \quad $[B_1]$\\
\quad $R_4 \rightarrow Z_0$ \quad $[A_2^{-1}]$\\
\quad $R_5 \rightarrow Z_1$ \quad $[A_2^{-1}]$\\
\quad $R_0 \rightarrow Z_2$ \quad $[A_2^{-1}]$
\end{tabular}
&
$10$ &
\begin{tabular}[t]{@{}l@{}}
Trans-CNOT: \\
\quad $X_0 \rightarrow R_2$ \quad $[B_2]$\\
\quad $X_1 \rightarrow R_3$ \quad $[B_2]$\\
\quad $X_2 \rightarrow R_4$ \quad $[B_2]$\\
\quad $R_5 \rightarrow Z_0$ \quad $[A_1^{-1}]$\\
\quad $R_0 \rightarrow Z_1$ \quad $[A_1^{-1}]$\\
\quad $R_1 \rightarrow Z_2$ \quad $[A_1^{-1}]$
\end{tabular}
&
$11$ &
\begin{tabular}[t]{@{}l@{}}
Trans-CNOT: \\
\quad $X_0 \rightarrow R_3$ \quad $[B_3]$\\
\quad $X_1 \rightarrow R_4$ \quad $[B_3]$\\
\quad $X_2 \rightarrow R_5$ \quad $[B_3]$\\
\quad $R_0 \rightarrow Z_0$ \quad $[A_0^{-1}]$\\
\quad $R_1 \rightarrow Z_1$ \quad $[A_0^{-1}]$\\
\quad $R_2 \rightarrow Z_2$ \quad $[A_0^{-1}]$
\end{tabular}
\\
\bottomrule
\end{tabular}

\end{table*}

\begin{figure*}[t]
\centering
\includegraphics[width=1\textwidth]{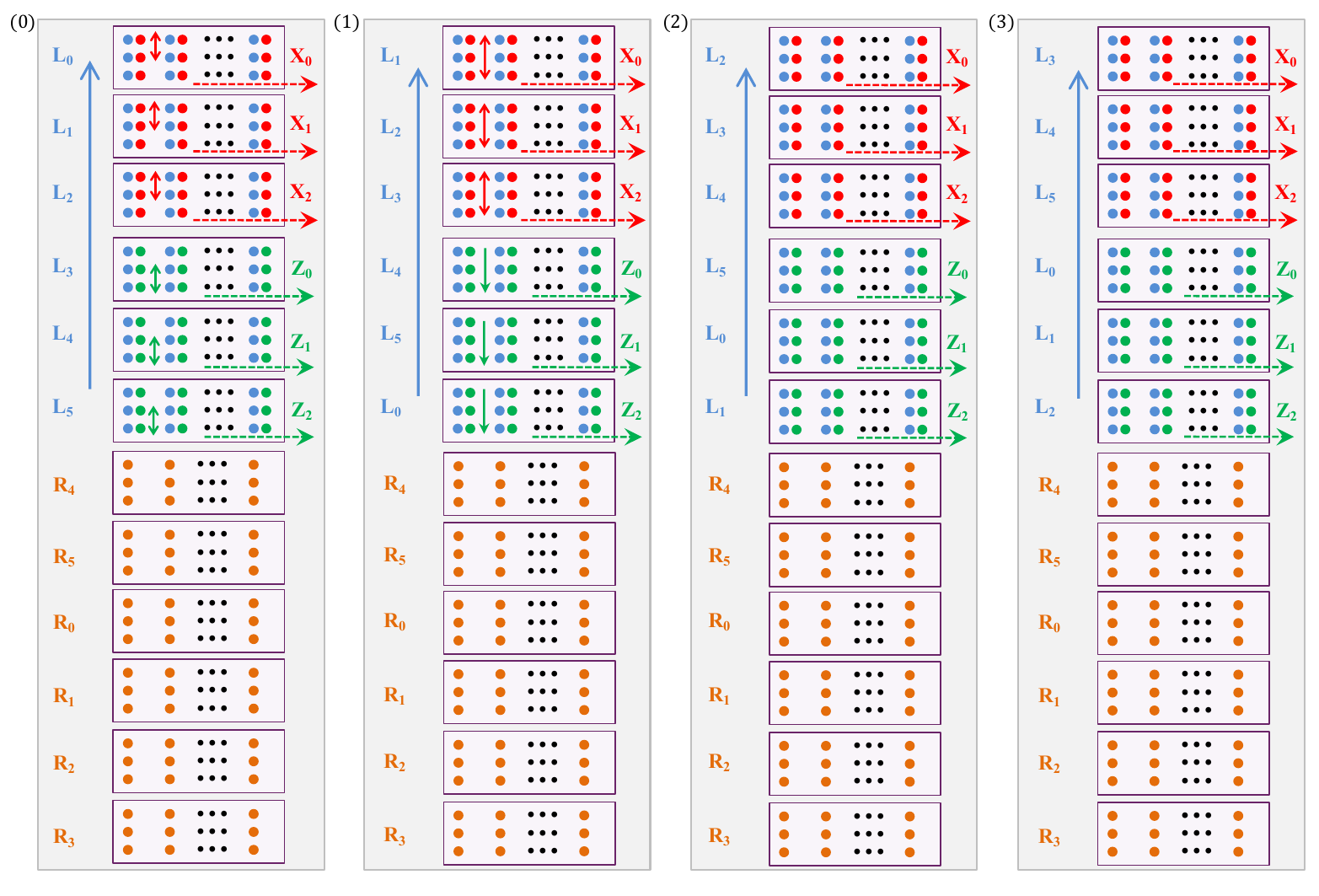}
\caption{\textbf{\boldmath{Atom rearrangement for one cycle of syndrome extraction for Cornucopia codes (steps $0-3$).}}
The remaining steps $4-11$ are detailed in Extended Data Fig.~\ref{fig:atom_moving2}. The atomic qubits comprising the $X$-, $Z$-, $L$-, and $R$-type blocks are denoted by red, green, blue, and orange circles, respectively. Data-block routing is indicated by vertical arrows: blue (yellow) one-way vertical arrows represent the block-level cyclic shifts of the six $L$-type ($R$-type) blocks. Black double-headed vertical arrows, which occur in steps $(5)$ and $(11)$, denote the macroscopic swapping of the top and bottom positions of the entire $L$ and $R$ block arrays. Ancilla row routing is indicated by solid vertical arrows: red (green) double-headed arrows indicate the swapping of two specific rows within the $X$-type ($Z$-type) blocks, whereas red (green) one-way vertical arrows represent the cyclic shift of all three rows within the $X$-type ($Z$-type) blocks. 
All these vertical movements (both data and ancilla routing) are deterministically fixed and identical across the entire Cornucopia code family. 
Finally, ancilla routing in the column direction is indicated by horizontal dashed arrows: red (green) one-way dashed arrows represent the bulk cyclic shifts of the $X$-type ($Z$-type) blocks in the column direction. Although the exact physical displacements of these column shifts depend on the parameters of the specific Cornucopia code, the shift distance remains uniform across all three sub-blocks of the same type ($X$ or $Z$) within any single step.}
\label{fig:atom_moving}
\end{figure*}

\begin{figure*}[t]
\centering
\includegraphics[width=0.98\textwidth]{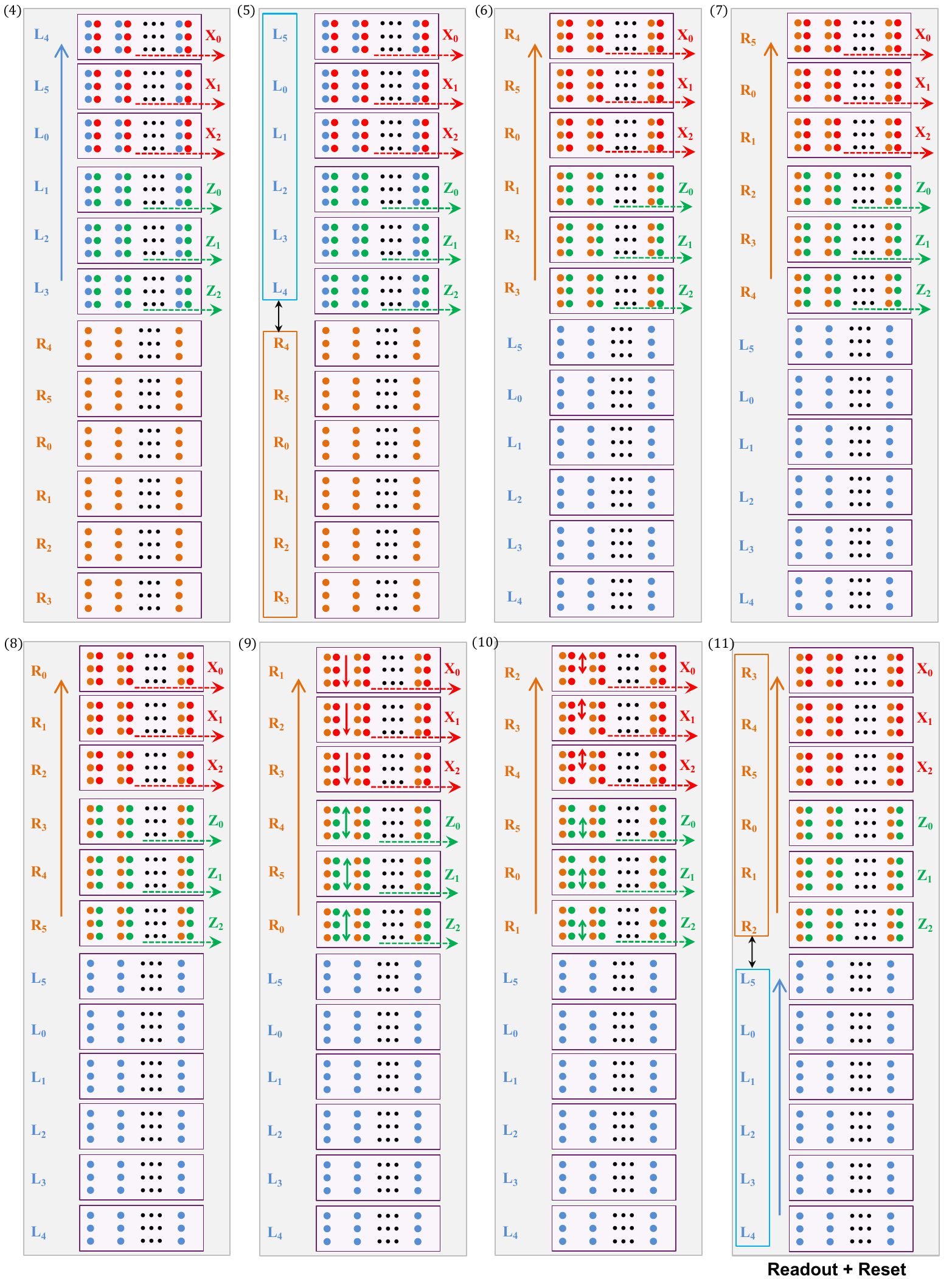}
\caption{\textbf{\boldmath{Atom rearrangement for one cycle of syndrome extraction for Cornucopia codes (steps $4-11$)}.} The notations follow those of Extended Data Fig.~\ref{fig:atom_moving}. 
The ``reset'' operation in neutral-atom arrays can be physically implemented by reloading fresh ancilla qubits.
}
\label{fig:atom_moving2}
\end{figure*}

\end{document}


\title{Supplementary Information for ``Quantum error correction at ultra-low overhead''}

\maketitle
\tableofcontents


\section{Cornucopia code construction}
\subsection{Proof of Lemma 1}
For convenience, we restate the lemma below.
Let $\operatorname{wt}(v)$ denote the Hamming weight, $\ker(M)$ denote the null space comprising all binary vectors $v$ such that $Mv = 0 \pmod 2$, and let $\operatorname{rs}(M)$ denote the row space spanned by the rows of a matrix $M$. 
\begin{lemma} A Cornucopia code offers exactly equal distance for $X$-type and $Z$-type errors, $d_X = d_Z = d$, where the respective distances are defined as
\begin{equation*}
\begin{aligned}
d_X &= \min\{\operatorname{wt}(v) : v \in \ker(H_Z) \setminus \operatorname{rs}(H_X)\}, \\
d_Z &= \min\{\operatorname{wt}(v) : v \in \ker(H_X) \setminus \operatorname{rs}(H_Z)\}.
\end{aligned}
\end{equation*}
\end{lemma}

\begin{proof}
To prove $d_X = d_Z$, we construct a bijective, weight-preserving linear map $\mathcal{T}: \mathbb{F}_2^{12P} \to \mathbb{F}_2^{12P}$ that maps $\ker(H_X) \to \ker(H_Z)$ and $\operatorname{rs}(H_X) \to \operatorname{rs}(H_Z)$.
First, we define an operator $\mathcal{I}: (x,y) \mapsto (2x+2, -y)$. 
All arithmetic on the coordinate pairs $(x,y)$ is implicitly evaluated modulo $3$ for the row indices and modulo $q$ for the column indices.
Note that $\mathcal{I}^2 = I$. For the three types of coordinate permutations listed in the Methods of the main text, a direct calculation gives $\mathcal{I} A_k \mathcal{I} = A_k^{-1}$ and $\mathcal{I} B_k \mathcal{I} = B_k^{-1}$.
Let a data vector be $v = (v_L, v_R)^\top$, comprising blocks $v_{L,j}$ and $v_{R,j}$ for $j \in \{0, \dots, 5\}$. We define the global data transformation $\mathcal{T}$ as:
\begin{equation}
(\mathcal{T}v)_{L, j} = \mathcal{I} v_{R, 2-j}, \qquad (\mathcal{T}v)_{R, j} = \mathcal{I} v_{L, 2-j},
\end{equation}
which swaps the $L$- and $R$-type blocks, reflects the block index $j \mapsto 2-j \pmod 6$, and applies $\mathcal{I}$. 
We also define an invertible check-space transformation $\mathcal{J}$ acting on a check vector $c$ (with blocks $c_i$ for $i \in \{0,1,2\}$) as $(\mathcal{J}c)_i = \mathcal{I} c_{2-i}$. This is a valid bijection because $i \mapsto 2-i \pmod 6$ maps the active set $\{0,1,2\}$ exactly to itself.

We evaluate the $i$-th check block row of $H_Z (\mathcal{T}v)$:
$$
(H_Z \mathcal{T}v)_i = \sum_{j=0}^5 B_{i-j}^{-1} (\mathcal{I} v_{R, 2-j}) + A_{i-j}^{-1} (\mathcal{I} v_{L, 2-j}).
$$
Using $M^\top = \mathcal{I} M \mathcal{I}$, we obtain:
$$
(H_Z \mathcal{T}v)_i = \mathcal{I} \left[ \sum_{j=0}^5 B_{i-j} v_{R, 2-j} + A_{i-j} v_{L, 2-j} \right].
$$
Setting $\ell = 2-j \pmod 6$, we get $i-j = \ell-(2-i)$:
$$
(H_Z \mathcal{T}v)_i = \mathcal{I} \left[ \sum_{\ell=0}^5 A_{\ell-(2-i)} v_{L, \ell} + B_{\ell-(2-i)} v_{R, \ell} \right] = \mathcal{I} (H_X v)_{2-i} = (\mathcal{J} H_X v)_i.
$$
This establishes the exact matrix identity $H_Z \mathcal{T} = \mathcal{J} H_X$. 

Because $\mathcal{J}$ is invertible, $H_X v = 0 \iff H_Z (\mathcal{T}v) = 0$, establishing $\mathcal{T}(\ker H_X) = \ker(H_Z)$. For the row spaces, multiplying the identity $H_Z \mathcal{T} = \mathcal{J} H_X$ on the right by $\mathcal{T}$ and using $\mathcal{T}^2 = I$ gives $H_Z = \mathcal{J} H_X \mathcal{T}$. Since $\mathcal{J}$ is invertible on the check space, $\operatorname{rs}(H_Z) = \operatorname{rs}(H_X \mathcal{T}) = \operatorname{rs}(H_X) \mathcal{T}$. This is equivalent to $\mathcal{T}(\operatorname{rs}(H_X)) = \operatorname{rs}(H_Z)$.

Finally, following the standard CSS convention, a $Z$-type logical error is an element of $\ker H_X\backslash \operatorname{rs}(H_Z)$. The bijection $\mathcal{T}$ maps any such $Z$-type candidate $v$ exactly to an element $\mathcal{T}v \in \ker H_Z$. Moreover, if $\mathcal{T} v \in \operatorname{rs}(H_X)$, then applying $\mathcal{T}$ again gives $v \in \operatorname{rs}(H_Z)$, contradicting the assumption that $v$ is nontrivial. Therefore, $\mathcal{T}v \in \ker H_Z\backslash \operatorname{rs}(H_X)$. Because $\mathcal{T}$ consists solely of coordinate permutations, it perfectly preserves the Hamming weight. Thus, any $Z$-type logical error of weight $w$ maps bijectively to an $X$-type logical error of the same weight, conclusively proving $d_Z = d_X$.
\end{proof}

\section{Syndrome measurement circuit}
\subsection{Proof of Proposition 1}
The syndrome extraction schedule shown in Fig.~1\textbf{c} explicitly preserves stabilizer determinism as stated by Proposition 1 in the Methods of the main text. 
For convenience we restate the Proposition below, followed by the rigorous proof.

\begin{proposition} 
For any Cornucopia code, the following syndrome extraction schedule, comprising twelve layers of non-overlapping CNOT gates:
\begin{equation*}
\begingroup
\setlength{\arraycolsep}{2pt}
\begin{array}{r|cccccc|cccccc}
\hline
t: 
& 0 & 1 & 2 & 3 & 4 & 5 
& 6 & 7 & 8 & 9 & 10 & 11 \\
\hline
X: 
& A_0 & A_1 & A_2 & A_3 & A_4 & A_5 
& B_4 & B_5 & B_0 & B_1 & B_2 & B_3 \\
Z: 
& B_3^{-1} & B_2^{-1} & B_1^{-1} & B_0^{-1} & B_5^{-1} & B_4^{-1} 
& A_5^{-1} & A_4^{-1} & A_3^{-1} & A_2^{-1} & A_1^{-1} & A_0^{-1} \\
\hline
\end{array}
\endgroup
\end{equation*}
avoids all intra-layer data block conflicts. 
Furthermore, the parity of the relative time ordering (e.g., $X$-before-$Z$) across all overlaps between any $X$-type and $Z$-type check is always even, guaranteeing deterministic stabilizer readouts.
\end{proposition}
\begin{proof}
The proof proceeds in two parts. We first verify the absence of data block conflicts, and subsequently prove the even timing parity of check overlaps.
\begin{itemize}
\item \textbf{Absence of data block conflicts.} 
Let $S=\{0,1,2\}$ denote the three check block row indices for both $H_X$ and $H_Z$. In the first half of the schedule (layers $t = 0, \dots, 5$), all $X$-type check blocks apply the coordinate permutations $A_t$, collectively targeting the left data blocks indexed by $S + t \pmod 6$. Simultaneously, all $Z$-type check blocks apply $B_{3-t}^{-1}$, targeting the left data blocks indexed by $S-(3-t) \equiv S + t - 3 \pmod 6$. To ensure no physical data block is targeted by both $X$ and $Z$ checks simultaneously, these two index sets must be disjoint: 
\begin{equation}
(S + t) \cap (S + t - 3) = \emptyset,
\end{equation}
for $t = 0, \dots, 5$. The specific set $S=\{0,1,2\}$ naturally satisfies this condition. In the second half of the schedule (layers $t = 6, \dots, 11$, with index $j = t - 6$), the $X$-checks apply $B_{j+4}$ and the $Z$-checks apply $A_{5-j}^{-1}$ on the right data blocks, collectively targeting the block sets $S + j+4 \pmod 6$ and $S + j - 5 \pmod 6$, respectively. This again necessitates $(S + j+4) \cap (S + j - 5) = \emptyset$, which is identically satisfied by $S=\{0,1,2\}$. Thus, the schedule enables parallel CNOT execution strictly without any intra-layer data qubit conflicts.
\item \textbf{Even timing parity of overlaps.} 
Consider a specific $X$-check $(r_X, u)$ and $Z$-check $(r_Z, v)$, where $r_X, r_Z \in \{0,1,2\}$ are the check block row indices and $u, v \in \mathbb{Z}_P$ are the local coordinates within the blocks. We define their block row difference as $\delta \equiv r_Z - r_X \pmod 6$. An overlap occurs if these checks act on the same physical data qubit, i.e., their support sets intersect.
For the $L$-type data blocks, the $X$-check acts via $A_i$ on the physical qubit $q_L = (r_X + i \pmod 6, A_i(u))$, while the $Z$-check acts via $B_j^{-1}$ on $q'_L = (r_Z - j \pmod 6, B_j^{-1}(v))$. Here, the physical qubits are indexed by the data block row index and local coordinates within the blocks.
An overlap ($q_L = q'_L$) requires the block constraint $r_X + i \equiv r_Z - j \pmod 6$, which simplifies to $i+j \equiv \delta \pmod 6$, and the local coordinate constraint $A_i(u) = B_j^{-1}(v)$, which is equivalent to $v = B_jA_i(u)$.
Similarly, for the $R$-type data blocks, the $X$-check acts via $B_j$ on $q_R = (6 + r_X + j \pmod 6, B_j(u))$, while the $Z$-check acts via $A_i^{-1}$ on $q'_R = (6 + r_Z - i \pmod 6, A_i^{-1}(v))$. Equating these yields the identical block constraint $i+j \equiv \delta \pmod 6$, alongside the local coordinate constraint $v = A_i B_j(u)$.

Crucially, we note that the only non-commuting operator pairs in our Cornucopia construction are $(A_1, B_2)$ and $(A_0, B_3)$, both of which satisfy $i+j=3$. However, the permissible block-row differences are strictly confined to $\delta \in \{0, 1, 2, 4, 5\} \pmod 6$. Since $3 \notin \{0, 1, 2, 4, 5\}$, all operator pairs $(A_i, B_j)$ that actually participate in active overlaps strictly pairwise commute ($A_i B_j = B_j A_i$). For these commuting pairs, the local coordinate conditions evaluate identically: $B_j A_i(u) = A_iB_j(u)$. 
As a result, any overlap occurring in the $L$-type data blocks is deterministically accompanied by a twin overlap in the $R$-type data blocks.

To guarantee deterministic measurement parity, we must ensure that the relative time ordering of the $X$-type and $Z$-type CNOT operations on the target data qubit is identical for both overlaps in this twin pair. From the explicit schedule, the exact execution layers are:
\begin{align}
t_X(A_i) &= i, \quad & t_Z(B^{-1}_j) &= [3-j]_6, \\
t_X(B_j) &= 6 + [j-4]_6, \quad & t_Z(A^{-1}_i) &= 11 - i,
\end{align}
where $[x]_6 \in \{0, 1, \dots, 5\}$ denotes $x \pmod 6$. We define $a = [3-j]_6$. The condition for an $X$-before-$Z$ ordering in the left overlap is $t_X(A_i) < t_Z(B_j)$, which directly simplifies to $i < a$.
Conversely, for the right twin overlap, the $X$-before-$Z$ condition is $t_X(B_j) < t_Z(A_i)$, yielding $6 + [j-4]_6 < 11 - i$. Because $a \equiv 3-j \pmod 6$, we have $j \equiv 3-a \pmod 6$. This implies $j-4 \equiv -1-a \equiv 5-a \pmod 6$. Since $a \in \{0, \dots, 5\}$, it follows that $5-a \in \{0, \dots, 5\}$, and therefore $[j-4]_6 = 5-a$. Substituting this exact identity gives:
$$6 + (5-a) < 11 - i \implies 11 - a < 11 - i \implies i < a.$$

The temporal conditions are mathematically identical: $t_X(A_i) < t_Z(B_j) \iff t_X(B_j) < t_Z(A_i)$. Therefore, the paired overlaps either both execute $X$-before-$Z$, or both execute $Z$-before-$X$. Their total contribution to the timing parity is always $0 \pmod 2$. 
\end{itemize}

This mechanism effectively pairs all overlapping CNOT operations generated by the commuting checks, forcing them to cancel out their parity contributions, thus ensuring deterministic stabilizer outcomes.
\end{proof}

\section{Logical routing}
\subsection{Proof of Proposition 2}
To formally characterize the logical action, let $T_c \colon (b,r,c) \mapsto (b,r,c+1)$ denote the global column-shift permutation, where $b$ is the data-block label, $r\in \mathbb Z_3$ is the row coordinate, and $c\in \mathbb Z_q$ is the column coordinate. We define the $X$- and $Z$-type logical Pauli quotient spaces as
\begin{equation*}
\mathcal L_X=\ker H_Z/\operatorname{row}(H_X),
\qquad
\mathcal L_Z=\ker H_X/\operatorname{row}(H_Z).
\end{equation*}
Since $T_c$ preserves the stabilizer group, it induces linear automorphisms $A_X\colon \mathcal L_X\to\mathcal L_X$ and $A_Z\colon \mathcal L_Z\to\mathcal L_Z$. The property $T_c^q=I$ naturally implies $A_X^q=I$ and $A_Z^q=I$. Letting $C_q=\langle u\mid u^q=1\rangle$ be the cyclic group generated by the column-shift action, we equip $\mathcal L_X$ and $\mathcal L_Z$ with $\mathbb F_2[C_q]$-module structures by defining the action of the generator $u$ as $A_X$ on $\mathcal L_X$ and as $A_Z$ on $\mathcal L_Z$. 
Equivalently, the group algebra is isomorphic to the polynomial quotient ring $\mathbb F_2[C_q]\simeq \mathbb F_2[u]/(u^q-1)$. As a module over itself, $\mathbb F_2[C_q]$ is the regular module of dimension $q$ over $\mathbb F_2$, where $u$ acts as a length-$q$ cyclic shift. 
Throughout, let $\mathbf{1}$ denote the one-dimensional trivial module (on which $u$ acts as the identity), and let $m\cdot M$ denote the direct sum of $m$ copies of a module $M$.

\begin{proposition}[Logical action of the column-shift automorphism]
For the Cornucopia code instances in Tab.~1 of the main text, all of which satisfy $k=n/2+4$, the induced action of $T_c$ on both logical spaces yields the module decomposition
\begin{equation}
\mathcal L_{X/Z}\simeq
18\cdot \mathbb F_2[C_q]\oplus 4\cdot \mathbf{1}.
\end{equation}
Equivalently, there exist symmetry-adapted bases for $\mathcal{L}_X$ and $\mathcal{L}_Z$ individually, such that within each space, a single physical column shift implements eighteen simultaneous logical cyclic shifts of length $q$, together with four fixed logical modes.
\end{proposition}

\begin{proof}
We prove the proposition by characterizing the $\mathbb F_2[C_q]$-module structure induced by the column-shift automorphism. All algebraic operations below, including ranks, kernels, and module decompositions, are performed over $\mathbb F_2$.

Let $R=\mathbb F_2[C_q]\simeq \mathbb F_2[u]/(u^q-1)$.
We consider the $X$-logical sector and introduce the chain complex
\begin{equation}
C_2\xrightarrow{d_2}C_1\xrightarrow{d_1}C_0,
\end{equation}
with $C_2\simeq R^9$, $C_1\simeq R^{36}$, and $C_0\simeq R^9$. 
For the $X$-logical sector, we identify $d_2=H_X^{\top}$ and $d_1=H_Z$, such that the homology is $H_1(C_\bullet)=\ker d_1/\operatorname{im}d_2 = \mathcal L_X$. The $Z$-logical sector is obtained by exchanging the roles of $X$ and $Z$.

We first evaluate the endpoint homologies $H_0(C_\bullet) = \operatorname{coker} d_1$ and $H_2(C_\bullet) = \ker d_2$. For the check matrices $H_X$ and $H_Z$ in Eq.~1 of the main text, there are three check block rows, with each block row containing $P=3q$ check rows. We define the rank deficiencies as
\begin{equation}
\epsilon_X = 9q - \operatorname{rank}_{\mathbb F_2}H_X,
\qquad
\epsilon_Z = 9q - \operatorname{rank}_{\mathbb F_2}H_Z.
\end{equation}
Because summing over the check block rows within the structural blocks introduces universal linear dependencies, we have $\epsilon_X, \epsilon_Z \ge 2$. The number of encoded logical qubits satisfies
\begin{equation}
k = 36q - \operatorname{rank}H_X - \operatorname{rank}H_Z = 18q + \epsilon_X + \epsilon_Z.
\end{equation}
For the Cornucopia code instances in Tab.~1 of the main text, $k = n/2 + 4 = 18q + 4$, which forces $\epsilon_X = \epsilon_Z = 2$. Hence, the endpoint homology modules have dimension two.
Moreover, the two universal dependency classes are explicitly generated
by differences of the three block-row sums, for example
\[
S_0+S_1,\qquad S_1+S_2 .
\]
Because a global column shift only permutes the columns within each
block, these classes are invariant under $T_c$. Hence, the induced
column-shift action on each endpoint space has two linearly independent
fixed vectors. Since the endpoint spaces are two-dimensional, the
action is exactly trivial:
\begin{equation}
H_0(C_\bullet)\simeq 2\cdot\mathbf 1,
\qquad
H_2(C_\bullet)\simeq 2\cdot\mathbf 1 .
\end{equation}

We next analyze the primary decomposition of the group algebra. Write $q = 2^s m$, where $m$ is odd, giving $u^q-1=(u^m-1)^{2^s}$. Since $m$ is odd, $u^m-1$ is square-free over $\mathbb F_2$ and can be factorized as $u^m-1=\prod_\alpha f_\alpha(u)$, where each $f_\alpha$ is irreducible. 
Hence, the group ring decomposes into primary components $R\simeq \bigoplus_\alpha R_{f_\alpha}$, where $R_{f_\alpha} = \mathbb F_2[u]/(f_\alpha^{2^s})$.
Consider any nontrivial primary component $f_\alpha(u)\neq u+1$. Because the endpoint homology modules are supported only on the invariant sector $u=1$, the localized endpoint homology vanishes: $H_0(C_\bullet)_{f_\alpha} = H_2(C_\bullet)_{f_\alpha} = 0$. Therefore, after localization at $f_\alpha$, the differential $d_1:R_{f_\alpha}^{36}\rightarrow R_{f_\alpha}^{9}$ is surjective and $d_2:R_{f_\alpha}^{9}\rightarrow R_{f_\alpha}^{36}$ is injective.
Reducing the localized complex modulo the maximal ideal $(f_\alpha)$, the middle homology dimension over the residue field is $36 - 9 - 9 = 18$. Thus, the $f_\alpha$-primary component contains exactly eighteen minimal generators. Furthermore, the Euler characteristic of the localized complex gives
\begin{equation}
\ell_{R_{f_\alpha}}\left(H_1(C_\bullet)_{f_\alpha}\right) = 18\ell(R_{f_\alpha}).
\end{equation}
A finite module over the local principal ideal ring $R_{f_\alpha}$ with eighteen generators and total length $18\ell(R_{f_\alpha})$ must be free of rank eighteen. Hence, for all $f_\alpha\neq u+1$,
\begin{equation}
H_1(C_\bullet)_{f_\alpha} \simeq R_{f_\alpha}^{18}.
\end{equation}
It remains to analyze the $u=1$ primary component. Let $x = u+1$, and let $A = \mathbb F_2[x]/(x^{2^s})$. The invariant component is therefore an $A$-module. The endpoint modules derived above become $H_0(C_\bullet)_{u+1} \simeq 2A/(x)$ and $H_2(C_\bullet)_{u+1} \simeq 2A/(x)$.

For odd $q$, we have $2^s=1$, and the invariant sector is semisimple.
In this case,
\[
A=\mathbb F_2[x]/(x)=\mathbb F_2 .
\]
The eighteen trivial components contained in the regular representation
together with the four additional invariant logical modes determined by
the endpoint structure give
\begin{equation}
H_1(C_\bullet)_{u+1}
\simeq
A^{18}\oplus4A/(x).
\end{equation}

For even $q$, the ring $A=\mathbb F_2[x]/(x^{2^s})$ is not semisimple, and endpoint homology alone does not uniquely
determine the extension structure of the middle homology. Therefore, we directly verify the module filtration of the induced logical action.
Let $N_{X/Z}=A_{X/Z}+I$. For the even-$q$ Cornucopia instances in Tab.~1, exact Gaussian elimination over $\mathbb F_2$ verifies
\begin{equation}
\dim\ker N_{X/Z}^{\,r} = 18r+4,
\qquad
1\le r\le 2^s .
\label{eq:even-q-filtration}
\end{equation}
For a nilpotent operator, the increment
\[
\dim\ker N^r-\dim\ker N^{r-1}
\]
equals the number of Jordan chains whose length is at least $r$. Therefore, Eq.~\eqref{eq:even-q-filtration} implies the presence of eighteen chains of maximal length $2^s$ and four chains of length one. Equivalently, the $u=1$ primary component has the decomposition
\[
A^{18}\oplus4A/(x).
\]
Thus,
\begin{equation}
H_1(C_\bullet)_{u+1}
\simeq
A^{18}\oplus4A/(x)
\end{equation}
for all even-$q$ instances considered here.
Combining all primary components, we obtain
\begin{equation}
H_1(C_\bullet)
\simeq
18\cdot R\oplus4\cdot\mathbf 1 .
\end{equation}

Thus, each of the logical spaces $\mathcal L_X$ and $\mathcal L_Z$ admits (independently) a symmetry-adapted decomposition of the form
\begin{equation}
\mathcal L_{X/Z} \simeq 18\cdot\mathbb F_2[C_q] \oplus 4\cdot\mathbf 1.
\end{equation}
This implies the existence of bases $\{\bar X_{\ell,t}\}$ and $\{\bar Z_{\ell,t}\}$, each adapted to the $C_q$ symmetry, in which the column-shift automorphism is represented block-diagonally as a cyclic shift within each sector:
\begin{equation}
\bar X_{\ell,t} \mapsto \bar X_{\ell,t+1}, \qquad
\bar Z_{\ell,t} \mapsto \bar Z_{\ell,t+1},
\end{equation}
where $\ell=1,\ldots,18$ and $t\in\mathbb Z_q$, supplemented by four invariant logical modes. Importantly, the bases for $\mathcal L_X$ and $\mathcal L_Z$ are symmetry-adapted independently, rather than canonically identified. In this representation, a single physical column shift acts simultaneously as a cyclic translation on each of the $18$ cyclic sectors.
\end{proof}

\section{Numerical simulation details}
Our numerical simulations employ a hierarchical, three-stage cascaded decoding strategy. This architecture consists of a primary relay-BP pass, a secondary relay-BP pass, and a final BP-OSD fallback stage.
The execution of the relay-BP decoder is governed by the following hyperparameters:
\begin{itemize}
\item $\gamma_0$: The memory strength applied to the error nodes of the first leg.
\item $[\gamma_i, \gamma_e]$: The sampling interval for the memory strengths applied across successive legs (fixed to $[-0.24, 0.6]$).
\item $T_m$: The maximum number of belief-propagation iterations allocated to the first leg.
\item $T^{\prime}_m$: The maximum number of belief-propagation iterations allocated to each subsequent leg.
\item $R$: The maximum number of sequential relay legs.
\item $S$: The target number of valid syndrome solutions required before termination, denoted as $S$.
\end{itemize}
The execution of the fallback BP-OSD decoder is governed by the maximum number of BP iterations $T_{\mathrm{osd}}$, the OSD processing method, and the min-sum scaling factor $s$. The decoder parameters and numerical simulation results for all considered Cornucopia codes, bivariate bicycle codes, and surface codes are detailed in Tab.~\ref{tab:decoding_statistics} and Tab.~\ref{tab:decoding_statistics_2}.

\begin{table*}[b]
\caption{
\textbf{Fitted coefficients for logical error rates.}
The logical error rate per logical qubit per cycle is fitted using
$p_L(p)=p^{d/2}\exp(c_0+c_1p+c_2p^2)$.
The fits for the Cornucopia codes use data with $p\leq0.0025$. For the bivariate bicycle and surface codes, the fits use data with $p\leq0.0045$. All fits are performed using unweighted least-squares regression in logarithmic space on the averaged $X$- and $Z$-basis logical error rates.
}
\label{tab:fit_coefficients}
\centering
\renewcommand{\arraystretch}{1.6}
\setlength{\tabcolsep}{13pt}

\begin{tabular}{@{}c c r r r@{}}
\toprule
{\bfseries Code family}
& {\bfseries\boldmath$[[n,k,d]]$}
& {\bfseries\boldmath$c_0$}
& {\bfseries\boldmath$c_1$}
& {\bfseries\boldmath$c_2$} \\
\midrule

\multirow{7}{*}{Cornucopia}
& $[[252,130,6]]$    & $7.46$  & $610.4$     & $41\,877$ \\
& $[[576,292,8]]$    & $9.07$  & $-1\,284.8$ & $1\,057\,824$ \\
& $[[900,454,10]]$   & $15.07$ & $-4\,169.3$ & $2\,031\,588$ \\
& $[[1044,526,12]]$  & $8.08$  & $6\,755.6$  & $-302\,774$ \\
& $[[1764,886,14]]$  & $16.86$ & $2\,268.7$  & $1\,023\,269$ \\
& $[[2304,1156,16]]$ & $11.72$ & $12\,575.6$ & $-1\,303\,808$ \\
& $[[2844,1426,18]]$ & $10.36$ & $18\,370.0$ & $-2\,448\,603$ \\
\midrule

\multirow{2}{*}{BB}
& $[[144,12,12]]$ & $17.79$ & $898.3$      & $-30\,973$ \\
& $[[288,12,18]]$ & $32.21$ & $-1\,979.4$ & $396\,017$ \\
\midrule

\multirow{3}{*}{Surface}
& $[[169,1,13]]$ & $23.50$ & $138.7$   & $56\,287$ \\
& $[[361,1,19]]$ & $32.36$ & $2\,774.9$ & $-345\,164$ \\
\bottomrule
\end{tabular}
\end{table*}

\begin{table*}[t]
\caption{
\textbf{Numerical simulation details for Cornucopia codes.} 
Each $[[n,k,d]]$ instance describes a code that encodes $k$ logical qubits with distance $d$ into $n$ data qubits. $N_{\rm c}$ denotes the truncated number of syndrome extraction cycles used for the simulation.
The tuples for the primary (RBP) and secondary (sRBP) relay-BP decoders are formatted as $(\gamma_0, T_m, R, T^{\prime}_m, S)$.
The fallback BP-OSD tuple is formatted as $(T_{\mathrm{osd}}, \text{method}, s)$.
The number of recorded logical failures out of total Monte Carlo samples is reported for both the $Z$ and $X$ bases across varying physical error rates $p$.
}
\label{tab:decoding_statistics}
\centering
\footnotesize
\renewcommand{\arraystretch}{1.5}
\setlength{\tabcolsep}{6pt}

\begin{tabular}{@{}c c l c c c@{}}
\toprule
\multirow{2}{*}{\bfseries\boldmath$[[n,k,d]]$}
& \multirow{2}{*}{\bfseries\boldmath$N_{\rm c}$}
& \multirow{2}{*}{\bfseries Decoder parameters}
& \multirow{2}{*}{\bfseries\boldmath$p$}
& \multicolumn{2}{c}{\bfseries Failures~/~Samples} \\
\cmidrule(lr){5-6}
& & & & {\bfseries\boldmath $Z$ basis} & {\bfseries\boldmath $X$ basis} \\
\midrule

\multirow{6}{*}{$[[252,130,6]]$}
& \multirow{6}{*}{$6$}
& \multirow{6}{*}{\shortstack[l]{
RBP: $(0.1,200,20,100,1)$\\
sRBP: $(0.1,500,200,200,1)$\\
BP-OSD: $(300,\texttt{OSD\_CS},0)$}}
& $0.001$  & $119/50\,000$   & $142/50\,000$ \\
& & & $0.0015$ & $642/50\,000$   & $607/50\,000$ \\
& & & $0.002$  & $2198/50\,000$  & $2085/50\,000$ \\
& & & $0.0025$ & $1181/10\,000$  & $1204/10\,000$ \\
& & & $0.003$  & $2571/10\,000$  & $2641/10\,000$ \\
& & & $0.004$  & $60/100$        & $61/100$ \\
\midrule

\multirow{6}{*}{$[[576,292,8]]$}
& \multirow{6}{*}{$6$}
& \multirow{6}{*}{\shortstack[l]{
RBP: $(0.1,200,20,100,1)$\\
sRBP: $(0.1,500,200,200,1)$\\
BP-OSD: $(300,\texttt{OSD\_CS},0)$}}
& $0.001$  & $8/500\,000$   & $4/500\,000$ \\
& & & $0.0015$ & $32/200\,000$  & $19/200\,000$ \\
& & & $0.002$  & $47/50\,000$   & $76/50\,000$ \\
& & & $0.0025$ & $38/2000$      & $34/2000$ \\
& & & $0.003$  & $108/1000$     & $99/1000$ \\
& & & $0.004$  & $67/100$       & $61/100$ \\
\midrule

\multirow{6}{*}{$[[900,454,10]]$}
& \multirow{6}{*}{$6$}
& \multirow{6}{*}{\shortstack[l]{
RBP: $(0.1,200,20,100,1)$\\
sRBP: $(0.1,500,200,200,1)$\\
BP-OSD: $(300,\texttt{OSD\_CS},0)$}}
& $0.001$  & $9/10\,000\,000$  & $13/10\,000\,000$ \\
& & & $0.0015$ & $18/1\,000\,000$  & $11/1\,000\,000$ \\
& & & $0.002$  & $9/50\,000$       & $14/50\,000$ \\
& & & $0.0025$ & $22/2000$         & $15/2000$ \\
& & & $0.003$  & $104/1000$        & $123/1000$ \\
& & & $0.004$  & $82/100$          & $85/100$ \\
\midrule

\multirow{6}{*}{$[[1044,526,12]]$}
& \multirow{6}{*}{$6$}
& \multirow{6}{*}{\shortstack[l]{
RBP: $(0.1,200,20,100,1)$\\
sRBP: $(0.1,500,200,200,1)$\\
BP-OSD: $(300,\texttt{OSD\_CS},0)$}}
& $0.0015$  & $3/2\,000\,000$   & $3/2\,000\,000$ \\
& & & $0.00175$ & $16/1\,000\,000$ & $14/1\,000\,000$ \\
& & & $0.002$   & $42/250\,000$    & $32/250\,000$ \\
& & & $0.0025$  & $17/2\,000$      & $15/2\,000$ \\
& & & $0.003$   & $98/1\,000$      & $114/1\,000$ \\
& & & $0.004$   & $84/100$         & $91/100$ \\
\midrule

\multirow{4}{*}{$[[1764,886,14]]$}
& \multirow{4}{*}{$7$}
& \multirow{4}{*}{\shortstack[l]{
RBP: $(0.1,200,20,100,1)$\\
sRBP: $(0.1,500,200,200,1)$\\
BP-OSD: $(300,\texttt{OSD\_CS},0)$}}
& $0.0015$  & $3/5\,000\,000$ & $4/5\,000\,000$ \\
& & & $0.00175$ & $8/1\,000\,000$ & $6/1\,000\,000$ \\
& & & $0.002$   & $23/200\,000$   & $18/200\,000$ \\
& & & $0.0025$  & $15/1\,000$     & $12/1\,000$ \\
\midrule

\multirow{6}{*}{$[[2304,1156,16]]$}
& \multirow{6}{*}{$8$}
& \multirow{6}{*}{\shortstack[l]{
RBP: $(0.1,200,20,100,1)$\\
sRBP: $(0.1,500,200,200,1)$\\
BP-OSD: $(300,\texttt{OSD\_CS},0)$}}
& $0.0015$  & $2/10\,000\,000$ & $3/10\,000\,000$ \\
& & & $0.00175$ & $11/2\,000\,000$ & $13/2\,000\,000$ \\
& & & $0.002$   & $30/200\,000$    & $27/200\,000$ \\
& & & $0.0025$  & $20/1000$        & $24/1000$ \\
& & & $0.003$   & $265/1000$       & $289/1000$ \\
& & & $0.004$   & $98/100$         & $198/200$ \\
\midrule

\multirow{4}{*}{$[[2844,1426,18]]$}
& \multirow{4}{*}{$8$}
& \multirow{4}{*}{\shortstack[l]{
RBP: $(0.1,200,20,100,1)$\\
sRBP: $(0,1000,200,400,1)$\\
BP-OSD: $(300,\texttt{OSD\_CS},0)$}}
& $0.00175$  & $5/2\,000\,000$ & $6/2\,000\,000$ \\
& & & $0.001875$ & $13/500\,000$   & $5/500\,000$ \\
& & & $0.002$    & $10/100\,000$   & $8/100\,000$ \\
& & & $0.0025$   & $26/1000$       & $28/1000$ \\
\bottomrule
\end{tabular}
\end{table*}

\begin{table*}[t]
\caption{
\textbf{Numerical simulation details for bivariate and surface codes.}
The tuples for the primary (RBP) and secondary (sRBP) relay-BP decoders
are formatted as $(\gamma_0,T_m,R,T'_m,S)$.
The BP-OSD tuple is formatted as
$(T_{\rm osd},\mathrm{method},s)$.
The recorded logical failures out of the total Monte Carlo samples are reported separately for the $Z$ and $X$ logical bases.
}
\label{tab:decoding_statistics_2}
\centering
\footnotesize
\renewcommand{\arraystretch}{1.5}
\setlength{\tabcolsep}{10pt}

\begin{tabular}{@{}c c l c c c@{}}
\toprule
\multirow{2}{*}{\bfseries\boldmath$[[n,k,d]]$}
& \multirow{2}{*}{\bfseries\boldmath$N_{\rm c}$}
& \multirow{2}{*}{\bfseries Decoder parameters}
& \multirow{2}{*}{\bfseries\boldmath$p$}
& \multicolumn{2}{c}{\bfseries Failures~/~Samples} \\
\cmidrule(lr){5-6}
& & & & {\bfseries\boldmath$Z$ basis}
        & {\bfseries\boldmath$X$ basis} \\
\midrule

\multicolumn{6}{c}{\bfseries bivariate codes} \\
\midrule

\multirow{4}{*}{$[[144,12,12]]$}
& \multirow{4}{*}{$6$}
& \multirow{4}{*}{\shortstack[l]{
RBP: $(0.1,200,20,100,1)$\\
sRBP: $(0.1,500,200,200,1)$\\
BP-OSD: $(300,\mathtt{OSD\_CS},0)$}}
& $0.00175$ & $16/30\,000\,000$ & $13/30\,000\,000$ \\
& & & $0.002$   & $11/10\,000\,000$ & $15/10\,000\,000$ \\
& & & $0.0025$  & $41/5\,000\,000$  & $32/5\,000\,000$ \\
& & & $0.003$   & $65/2\,000\,000$  & $60/2\,000\,000$ \\
\midrule

\multirow{4}{*}{$[[288,12,18]]$}
& \multirow{4}{*}{$8$}
& \multirow{4}{*}{\shortstack[l]{
RBP: $(0.1,200,20,100,1)$\\
sRBP: $(0,1000,200,400,1)$\\
BP-OSD: $(300,\mathtt{OSD\_CS},0)$}}
& $0.003$  & $2/100\,000\,000$ & $2/100\,000\,000$ \\
& & & $0.0035$ & $3/60\,000\,000$  & $4/60\,000\,000$ \\
& & & $0.004$  & $11/10\,000\,000$ & $5/10\,000\,000$ \\
& & & $0.0045$ & $6/2\,000\,000$   & $4/2\,000\,000$ \\
\midrule

\multicolumn{6}{c}{\bfseries Surface codes} \\
\midrule

\multirow{4}{*}{$[[169,1,13]]$}
& \multirow{4}{*}{$6$}
& \multirow{4}{*}{\shortstack[l]{
RBP: $(0.1,200,20,100,1)$\\
sRBP: $(0.1,500,200,200,1)$\\
BP-OSD: $(300,\mathtt{OSD\_CS},0)$}}
& $0.002$  & $7/15\,000\,000$ & $6/15\,000\,000$ \\
& & & $0.0025$ & $13/5\,000\,000$ & $15/5\,000\,000$ \\
& & & $0.003$  & $19/2\,000\,000$ & $14/2\,000\,000$ \\
& & & $0.0035$ & $28/1\,000\,000$ & $44/1\,000\,000$ \\
\midrule

\multirow{4}{*}{$[[361,1,19]]$}
& \multirow{4}{*}{$8$}
& \multirow{4}{*}{\shortstack[l]{
RBP: $(0.1,200,20,100,1)$\\
sRBP: $(0,1000,200,400,1)$\\
BP-OSD: $(300,\mathtt{OSD\_CS},0)$}}
& $0.003$  & $4/20\,000\,000$ & $3/20\,000\,000$ \\
& & & $0.0035$ & $10/10\,000\,000$ & $12/10\,000\,000$ \\
& & & $0.004$  & $41/10\,000\,000$ & $32/10\,000\,000$ \\
& & & $0.0045$ & $10/1\,000\,000$  & $13/1\,000\,000$ \\
\bottomrule
\end{tabular}
\end{table*}

\clearpage